\documentclass[aps,pra,superscriptaddress,amssymb,showpacs,twocolumn]{revtex4-1}
\usepackage{amsfonts} 
\usepackage{amsmath}
\usepackage{amssymb}
\usepackage{graphicx}
\usepackage{hyperref}
\usepackage{color}
\usepackage{natbib}  
\usepackage{bm} 
\usepackage[utf8]{inputenc}

\providecommand{\U}[1]{\protect\rule{.1in}{.1in}}

\newcommand{\vect}[1]{\bm{#1}} 
\newcommand{\ten}[1]{{#1}} 
\newcommand{\vk}{\boldsymbol{k}} 
\newcommand{\vR}{\boldsymbol{R}} 
\newcommand{\vrv}{\boldsymbol{r}} 
\newcommand{\vrvo}{\boldsymbol{r}^0} 
\newcommand{\ak}{a_{\boldsymbol{k}}} 
\newcommand{\wk}{\omega_{\boldsymbol{k}}} 
\newcommand{\adk}{a^{\dagger}_{\boldsymbol{k}}} 
\newcommand{\gk}{g_{\boldsymbol{k}}} 
\newcommand{\pk}{p_{\boldsymbol{k}}} 
\newcommand{\xk}{x_{\boldsymbol{k}}} 
\newcommand{\mk}{m_{\boldsymbol{k}}} 
\newcommand{\wc}{\omega_c} 
\newcommand{\Rr}{|\Delta \vrvo_{\lambda\mu}|} %
\newcommand{\Rra}{\Delta \vrvo_{\lambda\mu}} %
\newcommand{\A}{\mathcal{A}} %
\newcommand{\B}{\mathcal{B}} %
\newcommand{\C}{\mathcal{C}} %
\newcommand{\X}{\mathcal{X}} %
\newcommand{\Y}{\mathcal{Y}} %

\renewcommand{\Re}{\mathop{\mathrm{Re}}}
\renewcommand{\Im}{\mathop{\mathrm{Im}}}

\begin{document}

\title{Gaussian entanglement induced by an extended thermal environment}

\author{Antonio A. Valido}
\email{aavalido@ull.es}
\affiliation{Instituto Universitario de Estudios Avanzados (IUdEA), Universidad de La Laguna, La Laguna 38203 Spain}
\affiliation{Dpto.~F\'{\i}sica Fundamental II, Universidad de La Laguna, 38203 Spain}
\author{Daniel Alonso}
\email{dalonso@ull.es}
\affiliation{Instituto Universitario de Estudios Avanzados (IUdEA), Universidad de La Laguna, La Laguna 38203 Spain}
\affiliation{Dpto.~F\'{\i}sica Fundamental, Experimental, Electr\'{o}nica y
Sistemas, Universidad de La Laguna, La Laguna 38203 Spain}
\author{Sigmund Kohler}
\affiliation{Instituto de Ciencia de Materiales de Madrid, CSIC, Cantoblanco, 28049 Madrid, Spain }
\email{sigmund.kohler@icmm.csic.es}


\begin{abstract}
We study stationary entanglement among three harmonic oscillators which are
dipole coupled to a one-dimensional or a three-dimensional bosonic environment.
The analysis of the open-system dynamics is performed with generalized quantum
Langevin equations which we solve exactly in Fourier representation.  The focus
lies on Gaussian bipartite and tripartite entanglement induced by the highly
non-Markovian interaction mediated by the environment. This
environment-induced interaction represents an effective many-parties interaction with a
spatial long-range feature: a main finding is that the presence of a passive
oscillator is detrimental for the stationary two-mode entanglement. Furthermore,
our results strongly indicate that the environment-induced entanglement
mechanism corresponds to uncontrolled feedback which is predominantly coherent
at low temperatures and for moderate oscillator-environment coupling as compared
to the oscillator frequency.
\end{abstract}

\date{\today}

\maketitle
\section{Introduction}

Entanglement is a subtle feature of composite quantum systems, which is
invariant under local operations, i.e., operations that act solely upon one
constituent. Not considering protocols for entanglement swapping, entangling two
subsystem requires an interaction between them \cite{horodecki20091}. Such
interaction need not be direct, but may be mediated by a further quantum system
or even a heat bath, despite that environmental degrees of freedom generally
cause decoherence \cite{zurek2003} which is detrimental to entanglement. For
example, the interaction with a common heat bath can entangle two otherwise
uncoupled systems even in the weakly dissipative Markovian regime
\cite{braun20021,plenio20021,benatti20031,benatti20101} by making use of decoherence-free
subspaces that include entangled states
\cite{doll20061,doll20071,shiokawa20091,wolf20111,kajari20121} or by correlated
quantum noise that provides non-Markovian effects
\cite{horhammer20081,zell20091,vasile20101,fleming20121,correa20121}. Also more involved
system-environment interactions such as an exponential-like coupling
\cite{duarte20091,valente20101}, as well as dissipative engineering techniques
\cite{krauter20111} have been proposed for this issue. Given these
multi-faceted behavior, it is intriguing to investigate entanglement
between quantum systems in a more general dissipative scenario.

\begin{figure}[b]
\begin{center}
\includegraphics[width=5cm]{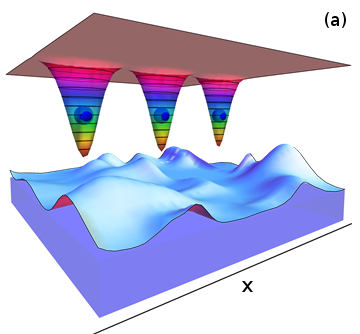} 
\includegraphics[width=6cm]{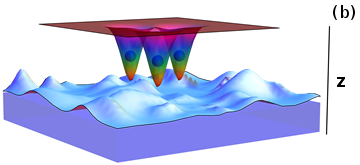}
\end{center}
\caption{\protect\label{figure21} Oscillator-environment configuration
considered in this work. Three oscillators are confined to the direction
indicated by the arrow; in the 1D arrangement (a) the $x$-direction, while in
the 3D configuration (b), the oscillators move only along the $z$-direction. The
interaction between the oscillators is mediated by a bosonic field, which also
causes decoherence and quantum dissipation.}
\end{figure}

In the present work, we investigate the setup sketched in Fig.~\ref{figure21}
and explore the influence of thermal relaxation on the
creation of stationary entanglement between three independent oscillators whose
equilibrium positions are spatially separated, such that the indirect
interaction mediated by the bath is retarded.  In particular we address two
issues.  The first one is the bath-induced entanglement formation between two
oscillators in the presence of a further oscillator.  The second one is the
characterization of the resulting stationary tripartite entanglement.  We
investigate both one-dimensional (1D) and three-dimensional (3D) environments,
where the former is restricted to a linear arrangement of the three oscillators.
Our model does not possess decoherence-free subspaces and, thus, any
emerging entanglement must stem from the environment-mediated interaction which
at the same time induces decoherence and quantum dissipation.
A most important feature of an extended environment is its dispersion relation which
implies a finite signal transmission velocity and, thus, causes retardation
effects. They may lead to an entanglement decay in several stages
\cite{doll20061,doll20071} or to a limiting distance for bath-induced two-mode
entanglement \cite{zell20091}.  Moreover, the dissipative quantum dynamics
acquires an additional non-Markovian influence, which in our case is rather
crucial because otherwise each oscillator would eventually reach its own Gibbs
state and, thus, the total state would be separable.

Our paper is organized as follows. In Sec.~\ref{model} we define our model and
derive within a quantum Langevin approach the main expressions and concepts used
later for the numerical computations, which tare presented and
discussed in Sec.~\ref{ther_ent}, where two-mode and three-mode entanglement is
studied as a function of the main parameters of the model. Conclusions are drawn
in Sec.~\ref{con}.  Some rather lengthy derivations have been deferred to the
appendix.

\section{The model system and equilibrium state}\label{model}

We employ a generalized Caldeira-Leggett model
\cite{weiss1999,leggett19871,hanggi19901} to capture thermal relaxation of the
oscillators, which can be derived from first principles
\cite{unruh19891,kohler20121}. We focus on the resulting stationary Gaussian
entanglement that stems from the quadratic form of the Hamiltonian. The
microscopic model will be approximately quadratic if the oscillators remain in
their equilibrium positions (which is compatible with the presence of the
environment-interaction effects), such that we can take the long-wave
approximation at lowest order.  The choice of a Gaussian initial state for the
reservoir guarantees the Gaussian nature of the final stationary state. We
assume a sudden switch-on of the interaction between the oscillators and the
bath, such that the initial state of the full system (oscillator modes plus
environment) is a product state $\rho_{0}=\rho \otimes \rho_{B}$.

In the case of a system composed by $N$ harmonic modes, the Gaussian stationary
state is determined up to irrelevant local displacements by the four $N \times
N$ correlation matrices
\begin{equation}\nonumber
C_{\vect A \vect B}(t-t')=\frac{1}{2}\langle \vect A(t) \vect B^T(t')+\vect B(t') \vect A^T(t)\rangle_{\rho},
\end{equation}
with $\vect A,\vect B \in \{\vect X,\vect P\}$ and where $\vect X$ and $\vect P$
denote column vectors with the position and momentum operators of the
oscillators. The Gaussian stationary state is characterized by the $2N \times
2N$  covariance matrix
\begin{equation}
\ten G=\begin{bmatrix}
      C_{\vect X \vect X }(0)&  C_{\vect X \vect P}(0) \\
      C_{\vect P \vect X }(0)&  C_{\vect P \vect P}(0)
      \end{bmatrix},
\label{cvmatrix}
\end{equation}
which contains the full information about the system fluctuations. To compute $
C_{\vect A \vect B}$, we employ the quantum Langevin equation formalism widely
used in the study of Brownian motion \cite{weiss1999,hanggi20051}, which we
adapt to our case of an extended environment. Regarding the study of
entanglement, remarkable achievements have reported been concerning its
classification and quantification for Gaussian states
\cite{giedke20011,adesso20071}.  A similar analysis of entanglement have been
recently carried out for three identical harmonic oscillators in an equilateral
triangular arrangement that are directly coupled and in contact with a common
bosonic field at zero temperature \cite{hsiang20131}. In the opposite scenario
of infinitely separated oscillators, each of them are fundamentally surrounded
by independent environments such that they might be at different temperatures.
In this case, the behavior of the entanglement has been proved to be essentially
different \cite{valido20131}.

\subsection{Generalized Langevin equation}\label{Langevin}

We consider three harmonic oscillators located at $\vR_{\lambda}=\vect
r^0_{\lambda}+\vect r_{\lambda}$, where $\lambda=1,2,3$, while $\vect
r^0_{\lambda}$ and $\vect r_{\lambda}$ denote equilibrium positions and
displacements, respectively. We attribute to each displacement a conjugate
momentum $\vect p_{\lambda}$, and employ the notations $\vect
r_{\lambda}:=(x_{\lambda},y_{\lambda},z_{\lambda})$ and $\vect
p_{\lambda}:=(p_{x,\lambda},p_{y,\lambda},p_{z,\lambda})$.  The oscillators are
assumed to be independent of each other with anisotropic confinement.  This
situation can be modeled by coupling the oscillators to a free bosonic field.
Following the above considerations, we model our setup by the system-bath
Hamiltonian $H_0 = H_S+H_B+ H_I$, with the system and the bath contribution
\begin{align}
H_S ={} &\sum_{\lambda=1}^3 \bigg(\frac{\vect p_{\lambda}^2}{2
m_{\lambda}}+\frac{1}{2}m_{\lambda}
(\omega^2_{x,\lambda}x^2_{\lambda}+\omega^2_{y,\lambda}y^2_{\lambda}+\omega^2_{z,\lambda}z^2_{\lambda})\bigg) ,
\\
H_B ={} & \sum_{\vk} \hbar \wk \adk \ak ,
\label{eq:1}
\end{align}
respectively, where
$\adk$ and $\ak$ are the usual
bosonic creation and annihilation operators for the bath mode with wavevector
$\vk = (2\pi/L) \mathbb{Z}^D$. We assume that only one degree of freedom per
oscillator is coupled to the bosonic field and, thus, experiences decoherence.
While in 1D, this assumption appears natural, it can be realized in the 3D case
by a strong anisotropy, $\omega_{x,\lambda} \ll
\omega_{y,\lambda},\omega_{z,\lambda}$, such that the motion in $y$- and
$z$-direction is frozen and can be ignored.  The interaction between the central
oscillators and the environment then takes the form
\begin{equation}
H_I=-\sum_{\lambda=1}^3 x_{\lambda} \sum_{\vk} \gk \Big( \ak e^{i \vk \cdot \vR_{\lambda}}+\adk e^{-i \vk \cdot \vR_{\lambda}} \Big),
\label{eq:2}
\end{equation}
with the coupling constants $\gk$ \cite{doll20061}. A technically
important simplification is provided by the assumption that
$e^{i \vk \cdot \vrv_{\lambda}} \ll 1$, which physically corresponds
to the long-wave limit or the dipole approximation for which we find
\begin{equation}
H_I\cong-\sum_{\lambda=1}^3 x_{\lambda} \sum_{\vk} \gk \Big( \ak e^{i \vk \cdot \vrvo_{\lambda}}+\adk e^{-i \vk \cdot \vrvo_{\lambda}} \Big).
\label{eq:3}
\end{equation}
When coupling the bosonic field to the oscillators a counter-term must be added if
one likes to preserve the bare oscillator potential of Eq.~\eqref{eq:1}.
Finally, the full oscillator-environment Hamiltonian becomes
$H_B+H_I \to H_{BI}$, where
\begin{equation}
\label{eq:5}
\begin{split}
 H_{BI}
={} & \sum_{\vk} \frac{1}{2\mk}\Big( \pk +\gk \sqrt{\frac{2 \mk }{\hbar \wk}} \sum_{\lambda=1}^3 x_{\lambda} \sin(\vk \cdot \vrvo_{\lambda})\Big)^2
\\ + &
\sum_{\vk} \frac{\mk \wk^2}{2} \Big( \xk-\frac{\gk}{\wk^2}\sqrt{\frac{ 2 \wk}{\mk \hbar}}\sum_{\lambda=1}^3 x_{\lambda} \cos(\vk \cdot \vrvo_{\lambda})\Big)^2.
\end{split}
\end{equation}
We have introduced the usual bosonic annihilation operator $\ak=(\mk \wk
\xk + i \, \pk)/\sqrt{2 \hbar \mk \wk}$ and its adjoint $\adk$. The coupling
together with the counter-terms in our Hamiltonian \eqref{eq:5} can be
interpreted as minimal coupling theory with $U(1)$ gauge symmetry
\cite{kohler20121}.  Moreover, in field theoretical terms, the oscillators are
coupled to the velocity of the bosonic field \cite{unruh19891}, which guarantees
that the energy remains positive definite and prevents ``runaway'' solutions
\cite{coleman19621}.

Associated to Hamiltonian (\ref{eq:5}) are equations of motion for
the degrees of freedoms of both the oscillators and the environment.
The dynamics of those of the oscillators, conditioned to the
environmental state, are given by a quantum Langevin equation which
follows from the exact Heisenberg equation of motion for $\vect
X:=(x_1,x_2,x_3)$ and which, after tracing out the environmental degrees of
freedoms, read \cite{weiss1999} (for details see Appendix \ref{App1})
\begin{equation}\label{GLE}
\ten{M} \ddot{\vect X}+ \ten \phi \vect X +
\frac{1}{\hbar}{\displaystyle\int_{-\infty}^{t}}
d\tau{\ten{\chi}}(t-\tau)\vect{X}(\tau)
=\vect{F}(t) ,
\end{equation}
where here the mass matrix $\ten M$ is proportional to the unit matrix, $\ten
M_{\lambda \mu}=m\delta_{\lambda \mu}$, while the counter-term
$\tilde{\Omega}_{\lambda \mu}$ is part of the potential matrix $\phi_{\lambda
\mu}=m\omega^2_{\lambda}\delta_{\lambda \mu}+2 \tilde{\Omega}_{\lambda \mu}$.
The memory-friction kernel $\ten\chi(t)$ has the form of a $3\times3$
matrix, and $\vect F$ is the column vector with the fluctuating forces
$F_{}(\vrvo_{\lambda},t):=F_{\lambda}(t)$ that act upon each oscillator. These
forces depend on the position of the oscillators and the environment. Owing to
their quantum nature, the forces are operators and commute with each other only
for time-like separations, i.e., $[F_{\lambda}(t'),F_{\mu}(t)]=0$ if
$|\vrvo_{\lambda}-\vrvo_{\mu}|> c|t-t'|$, where $c$ is the sound velocity of the
environment (or the speed of light, in a corresponding optical setup) which
enters via the dispersion relation $\wk = c|\vk|$.  It relates to the
memory-friction kernel via the Kubo formula
\begin{equation}
\label{chi}
{\ten \chi}_{\lambda\mu}(t-t^{\prime})
=-i \left\langle [ F_{\lambda}(t),F_{\mu}(t^{\prime})]\right\rangle_{\rho_{B}} \Theta( t-t^{\prime}-\vert \Delta \vrvo_{\lambda\mu}\vert/c),
\end{equation}
where the Heaviside step function $\Theta$ reflects causality with a retardation
stemming from the distance $\Delta \vrvo_{\lambda\mu}:=
\vrvo_{\lambda}-\vrvo_{\mu}$ between the oscillators $\lambda$ and $\mu$.  The
average has been taken with respect to the Gibbs state $\rho_B$ with temperature
$T$, which ensures the Gaussian property exploited below.  In the frequency
domain, the real part of the symmetrized forces correlation
$F_{\lambda}(t)F_{\mu}(t^{\prime})$ reads
\begin{equation}
\Re\left\langle {\vect F}(\omega)\vect F^{T}(\omega^{\prime})+{\vect F}(\omega^{\prime})\vect F^{T}(\omega)\right\rangle _{\rho_{B}}
= 4 \pi\hbar\delta\left(\omega+ \omega^{\prime}\right) \ten\Gamma (\omega)
\label{F-D}
\end{equation}
with the matrix $\Gamma$ defined by its elements
\begin{align}
\Gamma_{\lambda\mu}(\omega)
={}& -\frac{1}{\hbar}\Im \chi_{\lambda\mu}(\omega)
\coth \left(\frac{\hbar\omega}{2k_BT}\right) \nonumber
\\
={}& J_{\lambda,\mu}(|\omega|) \coth \left(\frac{\hbar|\omega|}{2k_BT}\right) .
\label{gamma}
\end{align}
This expression relates the real part of $\langle \left[
F_{\lambda}(t),F_{\mu}(t^{\prime}) \right] \rangle$ (commutator) to $\langle
\lbrace F_{\lambda}(t),F_{\mu}(t^{\prime})\rbrace \rangle$ (anti-commutator),
and thus, implies a quantum fluctuation-dissipation relation for the force
operators.  Moreover, we have introduced the bath spectral density
\begin{equation}
\label{J}
J_{\lambda,\mu}(\omega)= \frac{\pi}{\hbar} \sum_{\vk} g^{2}_{\vk}\cos\left( \vk \cdot \Delta \vrvo_{\lambda\mu}\right) 
\delta \left(\omega-\omega_{\vk} \right)
\end{equation}
which allows us to write the renormalization terms in the convenient form
\begin{eqnarray}
\tilde{\Omega}_{\lambda \lambda}&=&\frac{1}{\hbar}\sum_{\vk} \frac{g_{\vk}^{2}}{\omega_{\vk}}=\frac{1}{\pi }\int_{0}^{\infty}\frac{J_{\lambda,\lambda}(\omega)}{\omega}d\omega, \label{Omega}  \\
\tilde{\Omega}_{\lambda \mu}&=&\frac{1}{\hbar}\sum_{\vk} \frac{g_{\vk}^{2} }{\omega_{\vk}}\cos\left( \vk \cdot \Delta \vrvo_{\lambda\mu}\right) =\frac{1}{\pi}\int_{0}^{\infty}\frac{J_{\lambda,\mu}(\omega)}{\omega}d\omega.
\nonumber
\end{eqnarray}
With these relations, we can express the impact of the bath on the oscillators
and their effective interaction, as well as non-Markovian memory effects in
terms of the spectral density \eqref{J}.

The non-diagonal potential renormalization \eqref{Omega} couples the oscillator coordinates
$x_\lambda$ which, thus, are no longer the normal modes of our problem.  Therefore,
we introduce the transformation matrix $\ten{O}$ which maps to the normal modes
of the coupled oscillators, $\vect Q= \ten{O} \vect X $.  Together with the
according transformation for our matrices, we obtain for $\vect Q$ the Langevin
equation
\begin{equation}
\label{LangevinQ}
\ten{M} \ddot{\vect Q}+	\ten{\phi}_D \vect{Q}+
\frac{1}{\hbar}{\displaystyle\int_{-\infty}^{t}}
d\tau{\ten{\Xi}}(t-\tau)\vect{Q}(\tau)
=\vect{D}(t)
\end{equation}
with the mass matrix $\ten{O}\ten{M}\ten{O}^{T}=M$, the potential matrix
$\ten{\phi}_D=\ten{O}\ten{\phi}\ten{O}^{T}$, the susceptibility
$\ten{\Xi}(t)=\ten{O}\ten{\chi}(t) \ten{O}^{T}$, and the fluctuation forces
$\vect{D}(t)=\ten{O} \vect{F}(t)$, while the
fluctuation-dissipation relation becomes
\begin{equation}
\Re \left\langle {\vect D}(\omega)\vect D^{T}(\omega^{\prime})+{\vect D}(\omega^{\prime})\vect D^{T}(\omega)\right\rangle _{\rho_{B}}
= 4\pi\hbar\delta\left(\omega+ \omega^{\prime}\right) \ten\Upsilon (\omega) \nonumber
\end{equation}
with
$ 
\Upsilon(\omega)=-(1/\hbar)\Im \ten{\Xi}(\omega)=\ten{O}^{T}\Gamma\left( \omega\right)  \ten{O}.
$ 
While the conservative part of the transformed Langevin equation
\eqref{LangevinQ} is now diagonal, the modes may still couple via the
dissipation kernel $\ten{\Xi}(t)$, unless the latter is diagonal as well. This
can be achieved if $\ten{\phi}$ and $\ten{\chi}(t)$ commute at all times, which
is the case if all oscillators have the same fundamental frequencies and are
equally spaced, i.e., $\ten{\phi}$ and $\ten{\chi}(t)$ commute when the
equilibrium positions of the oscillators form a equilateral triangle ($\Delta
\vrvo_{\lambda\mu}=R$ for all $\lambda\neq\mu $) because they are symmetric
matrices and their product is also symmetric \cite{hsiang20131}. A further particular geometry is
given when the oscillators are placed in an isosceles triangle. Then the normal
mode corresponding to the relative motion of the oscillators placed at the ends
of the unequal side of the triangle and the center of mass dynamics are
independent of each other.  We consider these distinct geometries in
Sec.~\ref{ther_ent}.  Furthermore, it follows from the rank-nullity theorem
\cite{aitken1956} that the evolution of all normal modes will be subject to
dissipation and noise unless all oscillators have the same frequency and
are at the same place. Then their relative coordinate forms a decoherence-free
subspace \citep{shiokawa20091,kajari20121}.  In general however, i.e.,
for any other geometry, the oscillator-bath Hamiltonian does not possess a
decoherence-free subspace.

Having developed the formal solution of the quantum Langevin equation
\eqref{GLE}, we are able to evaluate the covariance matrix \eqref{cvmatrix}
whose entries read
\begin{align}
\label{cxx}
C_{\vect X \vect X} (0)={}& \hbar \int \frac{d\omega}{2 \pi} \ten \alpha(\omega) \ten \Gamma(\omega)\ten\alpha(-\omega)^{T},
\\
C_{\vect X \vect P}(0) ={}& C_{\vect P \vect X}(0)=
m \hbar{\displaystyle\int}
\frac{d\omega}{2 \pi}\, i \omega \,\ten \alpha(\omega) \ten \Gamma(\omega)\ten\alpha(-\omega)^{T}, \label{cxp}
\\
C_{\vect P \vect P}(0) ={}& m^{2} \hbar {\displaystyle\int}
\frac{d\omega}{2 \pi} \, \omega^2 \,\ten \alpha(\omega) \ten \Gamma(\omega)\ten\alpha(-\omega)^{T},\label{cpp}
\end{align}
where $\ten \alpha(\omega)$ corresponds to the Fourier transformed of the
left-hand side of the quantum Langevin equation \eqref{GLE}. All the covariances
contain the integration kernel $\ten{K}\left(\omega \right)=\ten \alpha(\omega)
\ten \Gamma(\omega)\ten\alpha(-\omega)^{T} $, while from the quantum
fluctuation-dissipation relation \eqref{F-D} follows that $\ten K\left(\omega
\right)$ is completely characterized by the generalized spectral density
$J_{\lambda,\mu}(\omega)$.


\subsection{Generalized spectral density and integration kernel $\ten {K}\left(\omega \right)$}
\label{uo}

We assume for the bosonic field the linear dispersion $\wk=c|\vk|$, which
comprises the physical cases of acoustic phonons and a free electromagnetic
field. Then it is possible to construct the spectral
densities $J_{\lambda, \mu}(\omega)$ which is a necessary step for computing the full
covariance matrix (\ref{cvmatrix}).  A detailed derivation for the expressions
introduced in this section can be found in Appendix~\ref{App2}.

We shall focus on 1D and 3D environments with isotropic coupling between the
oscillators and the bosonic field. For the coupling we choose $g^2_{\vk}=m \hbar
\gamma (\wk/\omega_c^{d-1})c^{d} V_{\vk}(d) e^{-\omega/\omega_c}$, where $d$ is
the dimension of the environment, $V_{\vk}$ is the number of field modes per
$d$-dimensional $\vk$-space volume, $\gamma$ is the coupling strength coupling,
and $\omega_c$ is the cut-off frequency of the environmental spectrum.
Eventually, the continuum limit $V_{\vk}\to 0$ will be taken. Hence, we obtain
the spectral densities
\begin{eqnarray}
J^{1D}_{\lambda ,\mu}(\omega)
&=& \pi m  \gamma \omega e^{-\omega/\wc} \cos(\omega \Rr/c),
\label{j1}
\\
J^{3D}_{\lambda, \mu}(\omega)
&=&\frac{4 \pi^2 m  c }{\Rr}\Big(\frac{\omega}{\wc}\Big)^2e^{-\omega/\wc} \sin(\omega \Rr/c).
\label{j3}
\end{eqnarray}
Accordingly, the potential renormalizations become
\begin{eqnarray}
\Omega^{1D}_{\lambda \mu}&=& \frac{ m \gamma \omega_{c}}{ 1+( \wc \Rr/c)^{2} },
\\
\Omega^{3D}_{\lambda \mu}&=& \frac{8 m \pi \gamma \omega_{c}}{[ 1+( \wc \Rr /c
)^{2}]^{2}}.
\end{eqnarray}
The imaginary part of the susceptibilities follows by inserting the spectral
densities into Eq.~\eqref{gamma}, while their real parts are conveniently obtained
via the Kramers-Kronig relations, so that we obtain
\begin{eqnarray}
\chi^{1D}_{\lambda\mu}(t)
&=& 4 m\, \gamma\, \hbar \, \omega^2_c \Theta\left( t- \Rr /c \right) 
\nonumber\\&&\times
\frac{\wc \Rr /c-t \wc }{[1+(\wc \Rr /c-t \wc)^2  ]^2 } ,
\label{chi1} 
\\
\chi^{3D}_{\lambda\mu}(t)
&=&8 \pi m \, \gamma\, \hbar \, \frac{\wc c}{\Rr}
\Theta( t- \Rr /c )
\nonumber\\ && \times \Bigg(  
\frac{1-3(\wc \Rr /c+t \wc)^2}{[1+(\wc \Rr /c+t \wc)^2  ]^3}
\nonumber\\ && -
\frac{1-3 (\wc \Rr /c-t \wc)^2}{[1+(\wc \Rr /c-t \wc)^2  ]^3}
\Bigg) .
\label{chi3}
\end{eqnarray}
The non-exponential decay in time obeyed by the susceptibilities (memory
kernels) describes non-Markovian dissipation \cite{zhang20121}, which will
turn out as essential ingredient for
stationary entanglement in our system. Moreover, the dimensionless
parameter $\Rr \wc /c$ is also involved in the renormalization terms and
generalized spectral densities. It compares two different time scales, on one
the hand $\Rr/c$, that is the time of flight of a phonon or photon between two oscillators,
and on the other hand $\wc^{-1}$ which represents the time scale during which memory
effects decay. Surprisingly, the environment-mediated interaction,
inherent in the susceptibilities and in the renormalization term, establishes an
effective coupling between all oscillators irrespective of their distance.  At
fixed time, they decay polynomially in space at least as $\sim ( \Rr \wc /c)^3 $
and $\sim ( \Rr \wc /c)^8 $ for the 1D and the 3D reservoir, respectively.
Although this interaction possesses long-range features, we shall see that the
characteristic length of the entanglement correlation is determined by $\Rr \wc
/c$, in agreement with Ref.\ \cite{zell20091}.

Once we have the susceptibilities and the renormalization terms at hand, we can
compute the matrices $\ten \alpha(\omega)$ for which we obtain
\begin{eqnarray}
\ten \alpha^{1D}_{\lambda\mu}(\omega)
&=& m(\omega_{\lambda}^{2}- \omega^{2}) 
    \delta_{\lambda\mu} -m \gamma \omega \Re\left( g(\omega)-g(-\omega)\right) \nonumber
\\ &&+
    \pi m \, \gamma \, \omega \, \Im\Big[\Theta(\omega ) 
e^{-\left(1/\omega_{c}-i \Rr/c \right)\omega}
\nonumber\\&& -
\Theta(-\omega) e^{(1/\omega_{c}-i \Rr/c)\omega} \Big] \nonumber \\
 &&-i \pi m \gamma \omega  \cos (\Rr \omega/c )
e^{-|\omega|/\wc} ,
\label{alphao}
\\
\ten \alpha^{3D}_{\lambda\mu}(\omega)
&=& m( \omega_{\lambda}^{2}- \omega^{2})
    \delta_{\lambda\mu}-i 4 \pi^2  m \gamma (c/\Rr) \Big(\frac{\omega}{\wc}\Big)^2
\nonumber\\&& \times \sin (\omega \Rr/c  )
e^{-|\omega|/\wc}
\nonumber \\
&&-\frac{4\pi \,m \, \gamma \,  c\, \omega^{2}}{\omega_{c}^{2}{\Rr}} \Im\left(g(\omega)+ g(-\omega)\right) \nonumber \\
&&- \frac{4\pi^2 \,m \, \gamma \,  c\, \omega^{2}}{\omega_{c}^{2}{\Rr}}
    \Re\Big[\Theta(\omega ) e^{-(1/\omega_{c}-i \Rr/c )\omega}
\nonumber\\&&
   +\Theta(-\omega) e^{(1/\omega_{c}-i \Rr/c )\omega} \Big] ,
\label{alphat}
\end{eqnarray}
where
\begin{equation}\nonumber
g(\omega)= e^{-(1-i \wc \Rr /c)\omega /\wc} \Gamma[0,-(1-i \wc \Rr /c)\omega /\wc],
\end{equation}
and $\Gamma(0,x)$ is the incomplete gamma function. With these expressions, we
readily obtain the elements of the stationary correlation matrix. Moreover, the
dimension-dependent integration kernels $K(\omega)$ can be evaluated to read
\begin{eqnarray}\nonumber
\ten{K}_{\eta,\beta}^{1D}(\omega)
&=& \pi m
\gamma \, \omega \, \coth\left(\frac{\hbar\omega}{2 k_{B}T}\right) e^{-|\omega|/\wc}
\\ && \times 
\sum_{\lambda,\mu}\cos\left(\omega \Rr /c \right)
\\ && \times 
\frac{\left( \hbox{adj}[\ten \alpha^{1D}(\omega)]\right)_{\eta\lambda} \left( \hbox{adj}[\ten \alpha^{1D}(-\omega)^{T}]\right)_{\mu\beta} }{|\ten \alpha^{1D}(\omega)||\ten \alpha^{1D}(-\omega)^{T}|}, \nonumber
\nonumber
\end{eqnarray}
\begin{equation}
\begin{split}
\ten{K}_{\eta,\beta}^{3D}(\omega)
={} & 4 \pi^2  m \gamma   \Big(\frac{\omega}{\wc}	\Big)^2
\coth\left(\frac{\hbar\omega}{2 k_{B}T}\right) 
e^{-|\omega|/\wc}
\\ & \times
\sum_{\lambda,\mu}\left( c/\Rr\right)\sin\left(\omega \Rr /c \right)
\\ & \times
\frac{\left( \hbox{adj}[\ten \alpha^{3D}(\omega)]\right)_{\eta\lambda} \left(
\hbox{adj}[\ten \alpha^{3D}(-\omega)^{T}]\right)_{\mu\beta}}{|\ten
\alpha^{3D}(\omega)||\ten \alpha^{3D}(-\omega)^{T}|} ,
\end{split}
\end{equation}
where $\hbox{adj} [ \ten \alpha ]$ and $|\ten \alpha|$ are the adjoint and the
determinant of $\ten \alpha$. With these expressions, we have achieved a closed,
albeit quite complicated form for the susceptibilities. Nevertheless, having the
analytic expressions at hand, certainly facilitates the numerical evaluation of
the covariance matrices \eqref{cxx}--\eqref{cpp}.


\section{Thermal entanglement induced by isotropic substrates}\label{ther_ent}

Having derived the solution of the quantum Langevin equations, we turn to the
entanglement among the oscillators induced by the non-Markovian dissipative
dynamics.  We focus on the quantum regime which requires low temperatures,
$k_{B} T \ll \hbar \omega_{\lambda}$.  In order to have the environment playing
a constructive role, it must couple strongly to the oscillators, such
that the quality factors $Q_\lambda = \omega_\lambda/\gamma \sim 1$--$10$ are
rather small. In this regime, the dissipative oscillator dynamics is strongly
non-Markovian.  In the numerical evaluations of our analytical expressions, we
use the typical units for nano oscillators, i.e., for masses $m=10^{-16}$\,kg,
for frequencies $\Omega=1 $\,GHz, and for distances $R=10$\,nm. Realistic values
for an environment realized by a solid-sate substrate are a cutoff frequency
(Debye frequency) corresponding to $\hbar\omega_c=6.58\cdot 10^{-2}$\,meV and
$c=3000$\,m/s for the speed of sound.

We characterize the Gaussian entanglement between two generic modes $\X$ and
$\Y$ by the logarithmic negativity \citep{vidal20021}
\begin{equation}
\label{LogNeg} 
E_{N}\left(\rho_{\X\Y} \right)
= \mathop{\text{max}}\lbrace0,-\ln\left( 2\nu_{-}\right) \rbrace,
\end{equation}
where $\X$ and $\Y$ represent one of the three oscillators, henceforth labeled
by $\A$, $\B$, and $\C$. Here, $\nu_{-}$ is the lowest symplectic eigenvalue of
the partial transpose covariance matrix $\ten G^ {T_{\Y}}$ (see Appendix
\ref{sectionppt} for details) corresponding to the reduced density matrix
$\rho_{\X\Y}$ of the two modes.  Regarding the analysis of three-mode Gaussian
entanglement, there is no generally accepted measure of tripartite entanglement
for arbitrary mixed states. Nonetheless it is possible to characterize it by a
classification scheme that assigns each state to one of five separability
classes \citep{giedke20011}, which range from fully inseparable states (class 1)
to mixed tripartite product states (class 5).  For details see
Appendix~\ref{sectionppt}.

Even though our focus lies on entanglement, we investigate for completeness also
the quantum fidelity $ \mathcal{F}\left(\rho,\rho^C \right)$ of the thermal
state $\rho^C \propto e^{-H_S/k_B T}$ as a function of the spatial degrees of
freedom and temperature.  In Ref.~\cite{marian20121} an analytical expression
for $\mathcal{F}\left(\rho,\rho' \right) $ was found for arbitrary $n$-mode
Gaussian states.  In our case, it becomes
\begin{equation}
\label{fidelity}
\begin{split}
\mathcal{F}\left(\rho,\rho^C \right)
={}& \prod_{i=1}^{n}\frac{2}{\left( \nu_{i}+\nu^C_{i}\right)^2 }
   \left[ \nu_{i}\nu^C_{i}+\frac{1}{4}\right.
\\ & 
   +\left.\sqrt{\left(\nu_{i}^2-\frac{1}{4} \right)
   \left(\left( \nu^C_{i}\right) ^2-\frac{1}{4} \right) } \right],
\end{split}
\end{equation}
where $\nu_{i}$ and $\nu^C_{i}$ are the symplectic eigenvalues of the covariance
matrix of $\rho$ and $\rho^C$, respectively.  Notice that here the symplectic
eigenvalues are different from those used for the logarithmic negativity,
because they are derived without partial transposition.

In previous works \cite{duarte20091,valente20101} on environment-induced
entanglement, it was found that when the oscillators are very close each other,
the most significant influence of the environment is to mediate an effective
interaction between the oscillators, while decoherence becomes relevant
mainly at higher temperatures. Moreover, it has been pointed out that for
identical oscillators, entanglement creation may stem from a decoherence-free
subspace \cite{shiokawa20091,kajari20121}.  Here, we consider oscillators with
different frequencies. Additionally, the Hamiltonian has no symmetries that
would support decoherence-free subspaces unless the distance between the
oscillators vanishes.  This implies that the stationary entanglement has its
roots in an environment-mediated interaction. From the Langevin
equation~\eqref{GLE}, we see that this interaction enters as a renormalization
potential or via dissipative effects, which we interpret as stochastic
feedback between the oscillators.
 
\subsection{Two-mode entanglement} \label{sec:2mode}

\begin{figure}[t]
\begin{center}
\includegraphics[width=6cm]{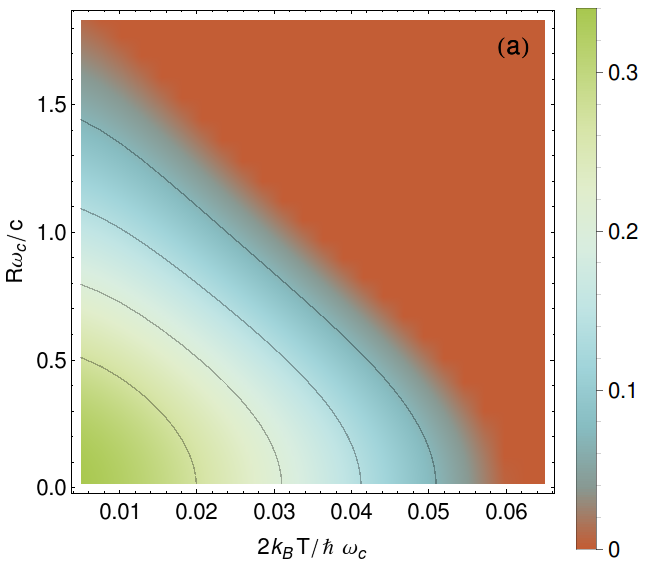} 
\includegraphics[width=6cm]{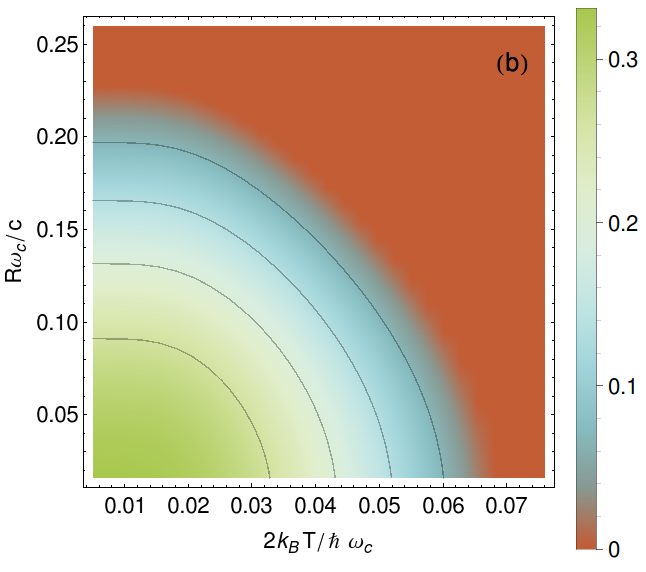} 
\end{center}
\caption{\protect\label{figure1}(Color online) Stationary two-mode entanglement
measured by the logarithmic negativity \eqref{LogNeg} as a function of
oscillator distance $R$ and temperature $T$ for a (a) 1D and a (b) 3D
environment. The oscillator frequencies are $\omega_{\A}=7.2\Omega$, and
$\omega_{\B}=13.2\Omega$, while the dissipation is $\gamma =5\Omega $.}
\end{figure}

We start by addressing the two-mode entanglement between the oscillators $\A$
and $\C$, placed at a distance $\Delta \vrvo_{\A\C}=R$, in the absence of
oscillator $\B$.  This is equivalent to putting oscillator $\B$ at infinite
distance, $\Delta \vrvo_{\A\B}=\Delta \vrvo_{\B\C}\rightarrow \infty $. Figure
\ref{figure1} depicts $E_{N}(\rho_{\A\C})$ for this case as function of the
distance $R$ and the temperature $T$ for a 1D and a 3D environment,
respectively.  Although the environment induces a long-range interaction [cf.\
the susceptibilities \eqref{alphao} and \eqref{alphat}] with a polynomial decay
in both space and time, we recover a central result of Ref.~\cite{zell20091}:
The correlation length is given by $R \approx \omega_c/c $, while the
entanglement vanishes at a finite distance $R_0$ which mainly depends on the
temperature, while being almost independent of the dissipation strength
$\gamma$.  Still a larger $\gamma$ supports the effective interaction required
for entanglement creation \cite{horhammer20081,duarte20091}, but also increases
decoherence which acts towards separability.  Nevertheless, as expected,
entanglement eventually disappears with increasing $\gamma$.

In 3D, entanglement generally appears more robust against thermal fluctuations,
which is consistent with previous findings for qubits
\cite{doll20061,doll20071}.  A qualitative explanation for this is the
super-Ohmic character of the 3D spectral density of the bath which leads to
stronger memory effects \cite{zhang20071}.  In turn, in the 1D case,
entanglement is less affected by increasing the spatial separation $R$, which
relates to the decay of the susceptibility as a function of the distance as we
mentioned above: As function of $R$, the susceptibility $\chi^{3D}(t)$ decreases,
at least, five orders stronger than $\chi^{1D}(t)$.  Thus, the effective
interaction at large distance in 3D is weaker than in 1D. In both cases, the
well-defined finite distance between the entangled oscillators
indicates that our mechanism for two-mode entanglement relies on memory effects.
Otherwise, we would expect a polynomial or exponential decay of the two-mode
correlations with increasing distance. This supports the idea that the
environment-induced interaction represents a kind of feedback between
oscillators which is predominantly coherent when only low energy environmental
modes are thermally excited, i.e., for $k_{B}T\ll\hbar \omega_c$.  Moreover,
depending on the separation, the coupling strength with the environment is not
too large to cause a strong decoherence.

We already discussed that the effective interaction potential provided by the
renormalization term $\tilde{\Omega}$ is crucial, but cannot explain fully the
amount of entanglement observed.  In order to underline this statement, let us
assume that dissipation and noise are negligible, so that the problem reduces to
two harmonic oscillators at thermal equilibrium with interaction potential $\ten
\phi$. Then identical oscillators with equal frequencies
$\omega_{\A}=\omega_{\C}=\Omega$ coupled at equal position to a substrate
($R\rightarrow 0$), will be entangled under a condition \cite{ludwig20101} that
in our case can be written as
\begin{eqnarray}\label{Ent1}
\left(2 N^{1D}_{+}+1 \right)\left(2 N^{1D}_{-}+1 \right)\left(1-\frac{2\gamma
\omega_c}{ \Omega^2 [ 1+( \wc R /c)^{2}]}\right) <1, \nonumber
\\
\\
\label{Ent2}
\left(2 N^{3D}_{+}+1 \right)\left(2 N^{3D}_{-}+1 \right)\left(1-\frac{16 \pi
\gamma \omega_c}{ \Omega^2 [ 1+( \wc R /c)^{2}]^{2}  }
\right)<1,\nonumber
\\
\end{eqnarray}
where
$N^{1D,3D}_{\pm}=[e^{\hbar\Omega^{1D,3D}_{\pm}/k_BT}-1 ]^{-1}$ denotes the bosonic
thermal occupation of normal modes with the frequencies 
\begin{eqnarray}
\Omega^{1D}_{\pm}&=&\Omega\sqrt{1+ \frac{2\gamma \omega_c}{ \Omega^2  }\pm
\frac{2\gamma \omega_c}{\Omega^2 [  1+( \wc R/c)^{2} ]}  }
\,,
\\
 \Omega^{3D}_{\pm}&=&\Omega\sqrt{1+ \frac{16 \pi \gamma \omega_c}{ \Omega^2
}\pm \frac{16 \pi \gamma \omega_c}{\Omega^2 [1+( \wc R/c)^{2}]^{2}}  }.
\end{eqnarray}
Notice that the conditions \eqref{Ent1} and \eqref{Ent2} result from an
expansion of the symplectic eigenvalues to first order in $\gamma \omega_c /\Omega^2$
implying $\gamma \omega_c < \Omega^2$, for which the left-hand side of these
expressions are strictly positive when neglecting dissipation and quantum noise
\cite{riseborough19851}. These conditions demonstrate that $R$ plays an
important role for the entanglement creation as can be appreciated in
Fig.~\ref{figure1}.  Still, these analytic considerations over-estimate the
influence of $R_0$ as a quantitative comparison with the numerically evaluated
expressions demonstrates (not shown). Although, we find that the available
entanglement generated by the effective potential $\tilde{\Omega}$ does not
display most of the characteristics of the stationary entanglement
discussed above, is still relevant in the transient dynamics
\cite{benatti20101}. In the long-time limit, both the numerical data and the
analytical results for the susceptibilities indicate that the mechanism
behind entanglement creation may be interpreted as uncontrolled feedback
(encoded by the susceptibility) which relies on the non-Markovian dissipation.

\subsection{Two-mode entanglement in the presence of a third oscillator}

\begin{figure}[t]
\includegraphics[width=6.5cm]{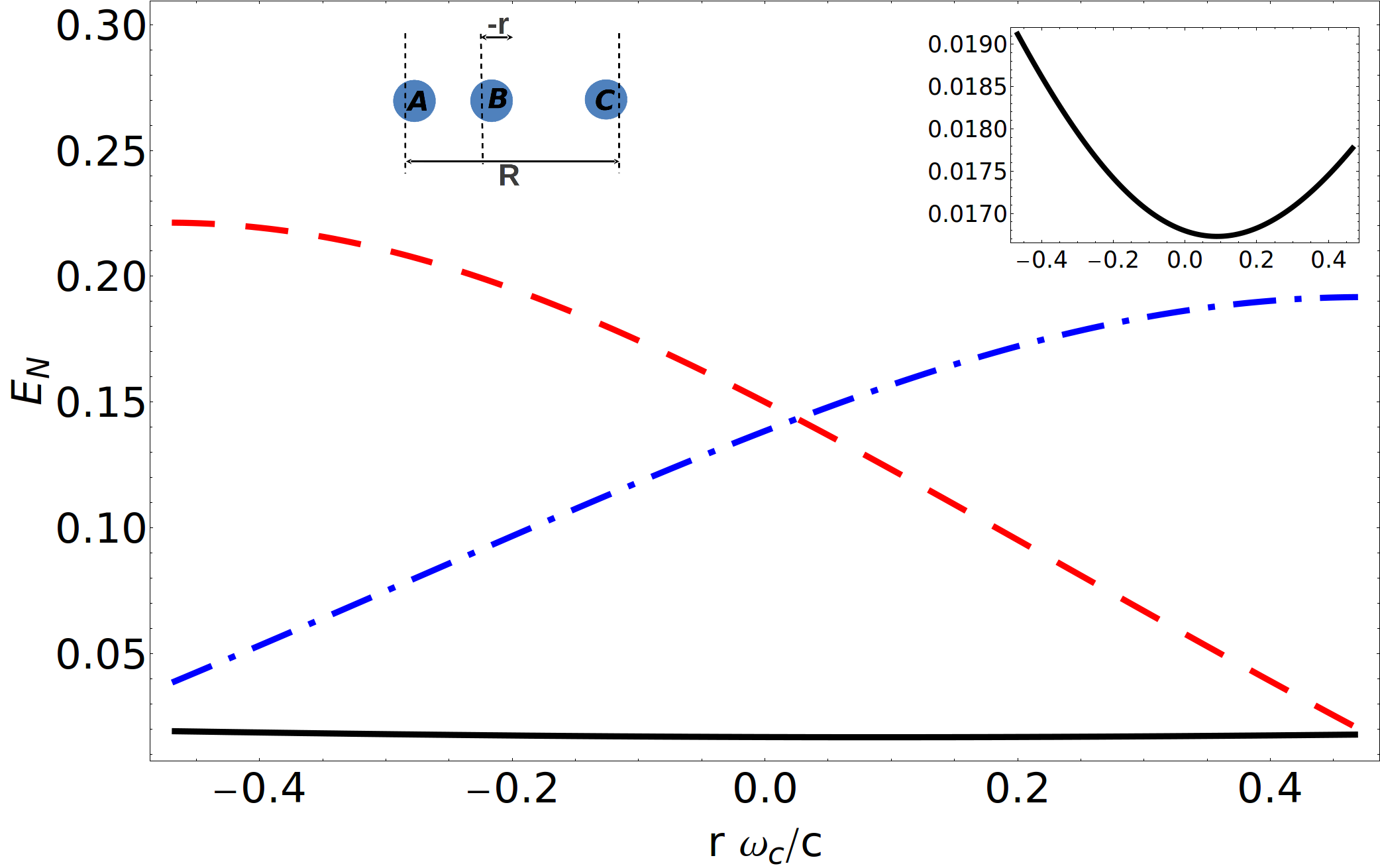} 
\caption{\protect\label{figure3}(Color online) Stationary two-mode entanglement
in the linear arrangement quantified by the logarithmic negativities
$E_N(\rho_{\A\C})$ (black solid line), $E_N(\rho_{\A\B})$ (red dashed), and
$E_{N}(\rho_{\B\C})$ (blue dash-dotted) for $R \omega_c/c = 0.933$.
Temperature and damping are $k_{B}T/\hbar \omega_c = 0.026$, $\gamma =
5\Omega $, respectively, while the frequencies are $\omega_{\A}=7.2\Omega$,
$\omega_{\B}=10.1\Omega$, and $\omega_{\C}=13.2\Omega$, where $\Omega= 1$\,GHz.
The asymmetry between $E_N(\rho_{\A\B})$ and $E_{N}(\rho_{\B\C})$ is a
consequence of choosing different oscillator frequencies.
The entanglement between $\A$ and $\B$ is less sensitive to a moderate increase
of temperature (not shown), because it involves the oscillators with the highest
frequencies. The inset is a zoom that demonstrates the small quadratic increase of
$E_N\left(\rho_{\A\C} \right)$.
}
\end{figure}

\begin{figure}[t]
\begin{center}
\includegraphics[width=7cm]{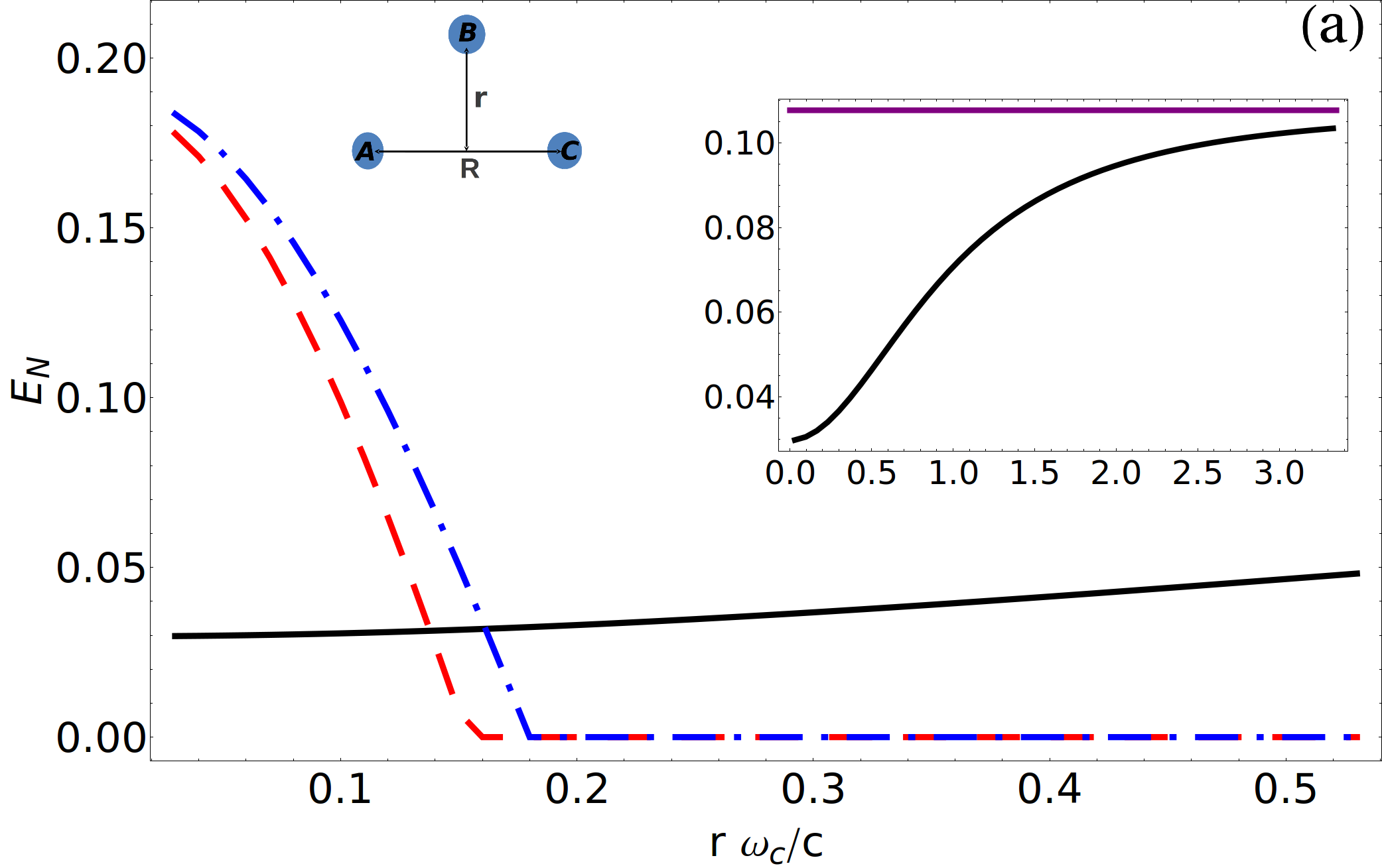} 
\includegraphics[width=6.9cm]{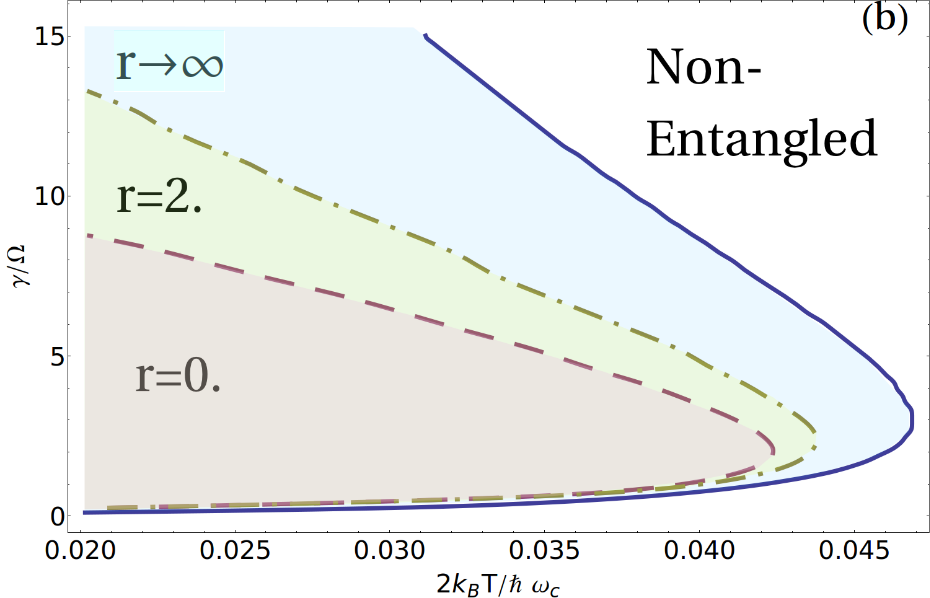} 
\end{center}
\caption{\protect\label{figure4}(Color online) (a) Stationary two-mode
entanglement measured by the logarithmic negativities $E_N(\rho_{\A\C} )$ (black
solid line), $E_N(\rho_{\A\B} )$ (red dashed line), and $E_{N}(\rho_{\B\C} ) $
(blue dash-dotted line) for the triangular geometry with $R \omega_c/c = 0.167$
as function of the displacement $r$.  All other parameters are as in
Fig.~\ref{figure3}.  The inset provides an extended picture of
$E_{N}(\rho_{\A\C} )$, where the purple flat line marks the value in the absence
of oscillator $\B$. (b) Phase diagram for fixed $R$ and various values of $r$ as
function of coupling strength $\gamma$ and temperature $T$.  In the shaded
areas, the oscillators $\A$ and $\B$ exhibit stationary entanglement.  The outer
blue line marks the limit $r\to\infty$, which is equivalent to the absence of
oscillator $\B$.  As oscillator $\B$ comes closer, the area with entanglement
shrinks.
}
\end{figure}

We have already seen that the coupling to a common environment induces an
effective interaction between oscillators and may create two-mode entanglement.
In the case that more oscillators are in contact with the bath, we expect that
additional effective interactions between any pair of oscillators emerge,
provided that the oscillators are sufficiently close to each other, i.e., for
distances $R\ll c/\omega_c$. It has been shown \cite{benatti20111,an20111} that
for three qubits in contact with a common environment, the two-qubit
entanglement for certain initial states persists in the long-time limit when
coupling a further qubit to the substrate.  Hence, the question arises how
two-mode entanglement is affected by the presence of a third oscillators.  We
study two different configurations: The first one is a linear arrangement in
which the three oscillators are coupled to a 1D environment with separations
$\Delta \vrvo_{\A\C}=R$, $\Delta \vrvo_{\A\B}=R/2+r$, and $\Delta
\vrvo_{\B\C}=R/2-r $, where $0<r < R/2$, as is sketched in Fig.~\ref{figure3}.
We fix $R$ such that the outer oscillators $\A$ and $\C$ may be entangled or
separable, depending on the other parameters. In the second configuration, the
oscillators are in contact with a 3D reservoir.  The oscillators $\A$ and $\C$
are again at distance $\Delta \vrvo_{\A\C}=R$, but oscillator $\B$ is shifted by
$r$ perpendicular to the line connecting $\A$ and $\C$, see sketch in
Fig.~\ref{figure4}.  Thus, $\Delta \vrvo_{\B\C}=\Delta \vrvo_{\A\B}=[r^2+(R/2)^2]^{1/2}$.

For the linear arrangement, we start by placing the oscillators $\A$ and $\C$ at a
distance $R$, and chose the other parameters such that both are separable in the
absence of oscillator $\B$, while for $r=0$, $\B$ is entangled with $\A$ in the
absence of $\C$ (and vice versa).  Then one might expect that the ``passive''
oscillator in the middle would give rise to an enhanced effective interaction
between $\A$ and $\C$, similar to what is found in harmonic chains with
nearest-neighbor interactions at thermal equilibrium \cite{anders20081}.
However, we find the opposite (not shown), namely that in the presence of
oscillator $\B$, one has to reduce the distance $R$ even below the limit found
above for the two-oscillator setup. Thus, the presence of oscillator $\B$ is
even harmful for the entanglement between the other two oscillators.  Therefore,
we chose for $R$ in the data shown in Fig.~\ref{figure3} a smaller value such
that $0<E_N(\rho_{\A\C})\ll1$.  As expected, oscillator $\B$ is stronger
entangled with the oscillator that is closer, which is in accordance with our
findings in the last section. The entanglement between the outer oscillators
stays rather small and remains almost unaffected by the position of the third
oscillator.  The small change can be appreciated in the inset of
Fig.~\ref{figure3}, which shows that $E_N(\rho_{\A\C})$ assumes its minimum when
$\B$ is roughly in the middle.

Our results for the triangular arrangement go into the same direction: We also
encounter that the third oscillator reduces the two-mode entanglement between
$\A$ and $\C$.  This generic behavior is in contrast to the one found
for setups that allow for decoherence-free subspaces \cite{benatti20111,an20111}.
The corresponding logarithmic negativity is plotted in Fig.~\ref{figure4}(a) as
function of the position of $\B$. In fact, the parameter space with
entangled states shrinks significantly by the presence of oscillator $\B$:
Fig.~\ref{figure4}(b) demonstrates that $E_N\left(\rho_{\A\C} \right)$ is
eventually destroyed when $\B$ is close enough to the pair. Then the
oscillator $\B$ becomes entangled with $\A$ and $\C$ almost simultaneously. That is,
$E_N(\rho_{\A\B})$ and $E_{N}(\rho_{\B\C})$ increase while
$E_N(\rho_{\A\C})$ becomes smaller. There is a trade-off between $E_N(\rho_{\A\C})$,
$E_N(\rho_{\A\B})$, and $E_{N}(\rho_{\B\C})$ resembling the monogamy property of
correlations \cite{horodecki20091}.
The competition between these three bipartite entanglements is characteristic
for our environment-induced entanglement mechanism, mainly because the
logarithmic negativity (i) is a bona fide measure that generally does not satisfy
monogamy and (ii) becomes increasingly manifest by raising the
coupling strength $\gamma$ as can be seen in Fig.~\ref{figure4}(b). This feature
is independent of whether the three oscillators have equal or different
frequencies. Furthermore, in the limit $r\to\infty$, $E_N(\rho_{\A\C})$
approaches the value of two-mode entanglement when the oscillator pair $\A\C$
evolves independent of $\B$. This shows that the oscillators effectively
interact even at distances greater than the correlation length of two-mode
entanglement, which implies that the environment-induced interaction has
long-range features.

Gathering the results for the two settings studied, they apparently
show that the environment-mediated interaction induces a trade-off between the
three two-mode entanglements. This feature is highly emphasized in the
triangular setting, where $\B$ is brought closer to both $\A$ and
$\C$. For identical oscillators, we observe that all possible two-mode
entanglements take the same values when they form an equilateral triangle, i.e.,
for $r=\sqrt{3}R/2$. At smaller values for $r$, the entanglements
$E_N(\rho_{\A\B})$ and $E_{N}(\rho_{\B\C})$ are larger than $E_N(\rho_{\A\C})$,
because $\A$ and $\C$ are further separated to each other than to $\B$.  One of
our main findings is that the presence of the oscillator $\B$ reduces the
entanglement between $\A$ and $\C$.  This tendency towards separability might be
enhanced by adding further oscillators. However even though $E_N(\rho_{\A\C})$
may be reduced or vanish in the presence of oscillator $\B$, there is still the
possibility of an emerging multi-partite entangled such as the formation of
GHZ-like states.  This emergence of tripartite entanglement on the expense of
smaller bipartite entanglement may be interpreted as consequence of an effective
three-body interaction by which all three oscillators act simultaneously via the
same bath.

\subsection{Three-mode entanglement}

For the characterization of multi-partite entanglement, we employ the
classification scheme for tripartite Gaussian entanglement developed by Giedke
\textit{et al.} \cite{giedke20011} and summarized in Appendix \ref{sectionppt}.
According to this scheme, each state falls in one of the following five
classes. C1: fully inseparable states, C2: two-mode biseparable states, C3:
two-mode biseparable states, C4: bound tripartite entangled states, and C5:
fully separable states.  Notice that class C1 is not a strict classification
but rather subsumes all so-called genuinely tripartite-entangled states
\cite{bennet20111}.

\begin{figure}[t]
\includegraphics[width=6cm]{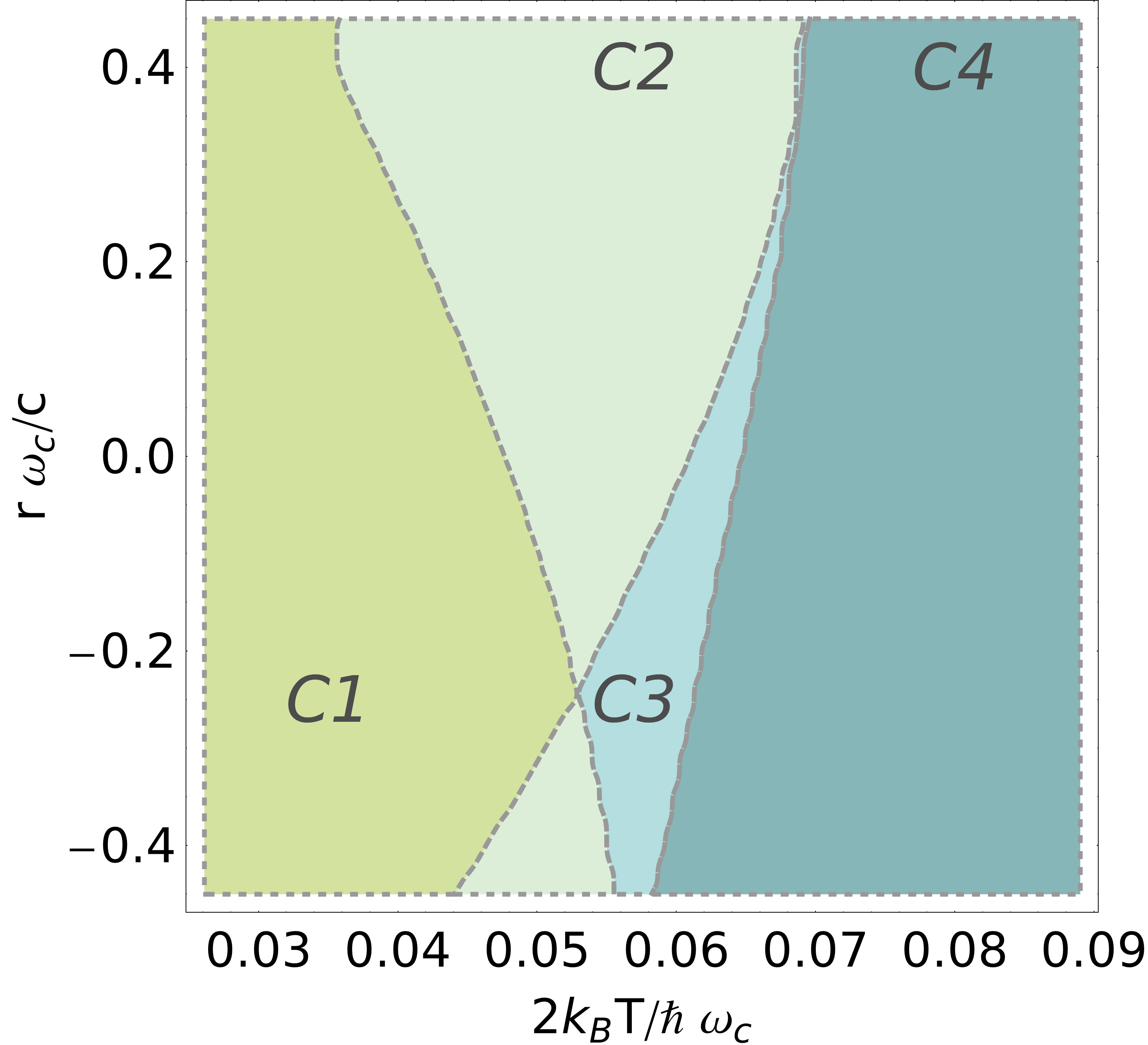} 
\caption{\protect\label{figure6}(Color online) Phase diagram of the separability
classes for the linear configuration as function of temperature and the position
$r$ of oscillator $\B$.  All other parameters are as in Fig.~\ref{figure3}.}
\end{figure}

Concerning tripartite entanglement a most important question is whether an
optimal arrangement for genuine tripartite entanglement exits.  The results of
the previous section suggest that equally spaced oscillators might be rather
unfavorable for two-mode entanglement (see inset in Fig.~\ref{figure3}). An
expectation inferred from those results (see Fig.~\ref{figure1}) is that
tripartite entanglement decreases with distance as bipartite entanglement does,
i.e., it should vanish at large distances.  Still it is interesting to see
whether three-mode entanglement is more robust against variation of $r$ than
two-mode entanglement.  Moreover, the limiting distance may be different from
$R\omega_c/c$.

\subsubsection{Linear arrangement}

\begin{figure}[t]
\begin{center}
\includegraphics[width=5cm]{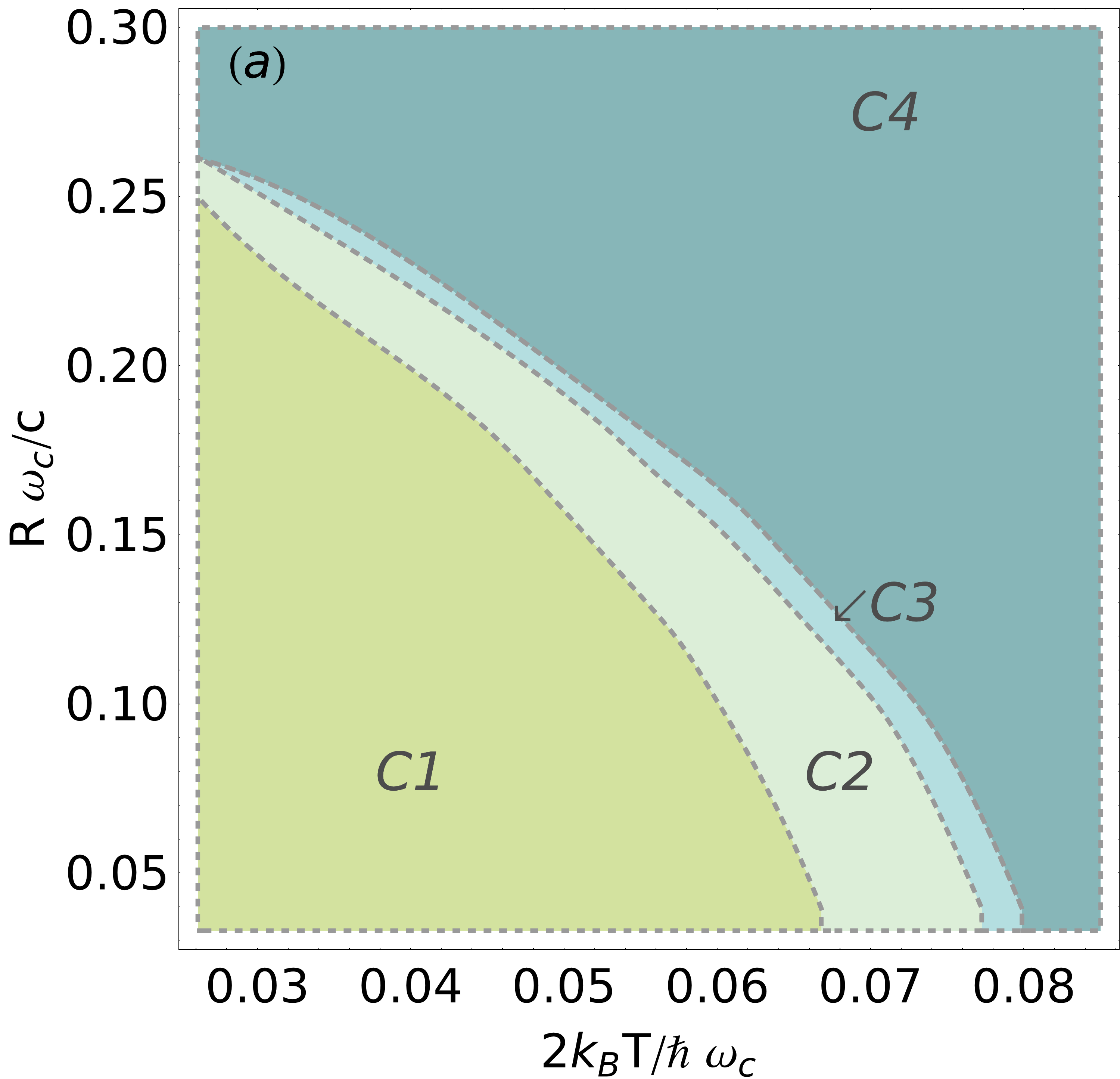} 
\includegraphics[width=7.1cm]{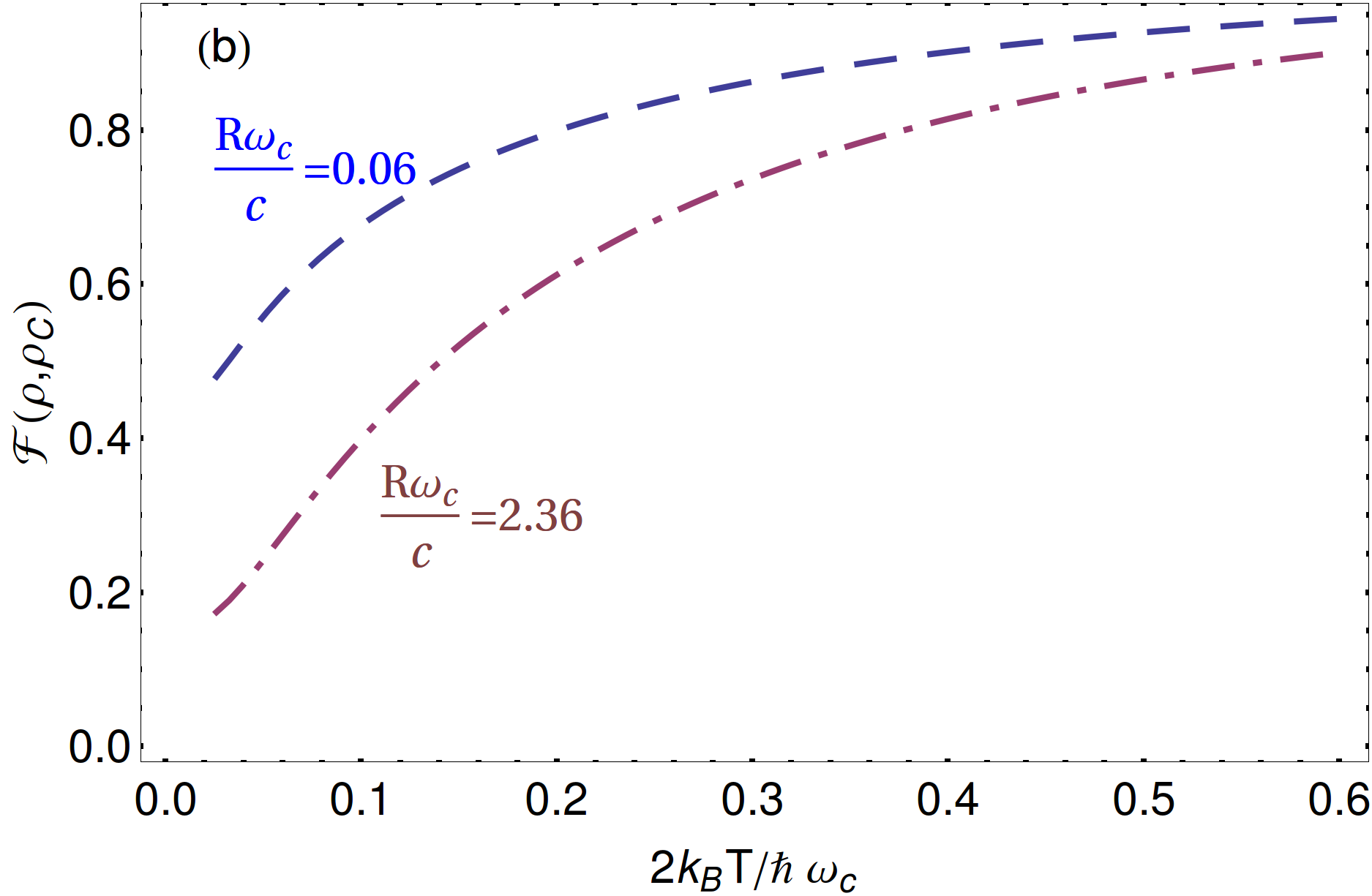} 
\end{center}
\caption{\protect\label{figure7}(Color online)
(a) Separability phase diagram for the equilateral-triangular for the oscillator
frequencies and coupling strengths used in Fig.~\ref{figure4}.
(b) Quantum fidelity between the stationary state and the thermal canonical
state as a function of the temperature for the distances $R \omega_c/c = 0.066$
(blue-dashed line) and $R \omega_c/c = 2.367$ (pink-dot-dashed line).
}
\end{figure}

Figure~\ref{figure6} shows the phase diagram of the separability classes for the
case in which all oscillators are coupled to a one-dimensional environment.
Most importantly, it demonstrates the relative robustness of the fully
inseparable states (class C1) against shifting the position of oscillator $\B$
and against moderate temperature increase.  Fully inseparable states are found
for small temperatures and when oscillator $\B$ is a bit closer to $\A$ than to
$\C$. This asymmetry stems from the fact that the oscillator $\C$ is less
affected by the thermal fluctuations than the other two oscillators, owing to
its larger frequency.  In general, we expect the genuine tripartite
entanglement to be rather insensitive to variations of the geometry as long as
all oscillators interact strongly in the same manner through the reservoir,
i.e., when oscillator $\B$ is roughly in the middle.  Otherwise, the geometry
could enhance the interaction between two particular oscillators, which may lead
to a situation in which the third oscillator becomes separable.  In the phase
diagram (Fig.~\ref{figure6}), this is visible in the emergence of regions with
separability class C2 when $r$ tends towards $\pm R/2$. Thus, in contrast to the
two-mode case, the equidistant placement of oscillator $\B$ at $r=0$ is the
optimal setting for genuine tripartite entanglement, at least in the case of
equal oscillators.

\subsubsection{Arrangement in an equilateral triangle}

Having noticed that in the 1D case, optimal tripartite entanglement is achieved
in the most symmetric situation, we restrict ourselves in the 3D case to the
configuration in an equilateral triangle with lateral length $R = \Delta
\vrvo_{\A\C}=\Delta \vrvo_{\B\C}=\Delta \vrvo_{\A\B}$.  Figure~\ref{figure7}(a)
depicts the corresponding separability phase diagram.  Again we find for small
$R$ and low temperatures that the stationary state is fully inseparable (class
C1).  With increasing temperature, we notice a transition via the one-, two-, and
three-mode biseparable classes C2, C3, and C4 to the fully separable class C5 at
high temperatures $T\gtrsim \hbar\omega_c/k_B$ (the latter is beyond the
plotted range).  The appearance of classes C2 and C3 obviously requires some
asymmetry in the setup, which stems from choosing different oscillator
frequencies.
In comparison to the two-mode entanglement studied in Sec.~\ref{sec:2mode},
however, tripartite bound entanglement (class C4) is more robust against
separation and temperature effects than for two modes.  Indeed we find that it
may survive up to values of $R\omega_c/c $ that clearly exceed unity.  This can
be explained by the fact that the susceptibilities reflect an effective coupling
of all oscillators independent of their spatial separation (cf.\ discussion in
the Sec.~\ref{uo}), which enables large-distance entanglement.  The latter is
also in agreement with the two-mode entanglement $E_N(\rho_{\A\C})$ discussed
above: It asymptotically approaches the value found for the oscillator pair
$\A\C$ in the absence of a third oscillator (see inset of Fig.~\ref{figure4})
and underlines that the environment induces long-range interaction. On the other
hand, the quantum fidelity \eqref{fidelity}, which shows the ``sophistication''
of the stationary state, reveals that the (fully separable) thermal state is
reached for $ k_{B}T/\hbar \omega_{c}\gtrsim 1$ [see Fig.~\ref{figure7}(b)],
irrespective of the distances between the oscillators.  Hence, only at high
temperatures, decoherence dominates so that here the full separability turns out
to be a decoherence phenomenon.

\section{Summary and Conclusions}\label{con}

We have studied the dynamics of three harmonic oscillators as a generic
tripartite system that becomes entangled through the interaction with a common
extended environment. The oscillators are embedded in a thermal bosonic heat
bath which we eliminated to obtain generalized quantum Langevin equations.
Although the oscillators are not directly coupled, the contact via the heat bath
provides an environment-mediated interaction which can induce bipartite and
tripartite entanglement among the oscillators. The equations of motion for a 1D
and 3D isotropic environment contain this interaction as a long-range coupling
entering via a renormalization term and through the susceptibility, which takes
the backaction into account. For both two-mode entangled and fully inseparable
oscillators, the characteristic correlation length is roughly given by the ratio
$R\omega_c/c$.  For a 3D environment it is smaller than for the 1D case.
Nevertheless, the entanglement generated by a 3D environment is more robust
against thermal fluctuations.  As expected, non-Markovian memory effects play a
crucial role for the dynamics.  Interestingly enough, there is a trade-off in
the attainable two-mode entanglement between the different oscillator pairs,
because the presence of a passive oscillator is detrimental for two-mode
entanglement. This provides strong evidence that the environment-induced
interaction is mainly a many-parties interaction that tends to favor
multi-partite correlations (here tripartite instead of bipartite), such that
GHZ-like states emerge. In general, the numerical data suggest that the
mechanism is based on uncontrolled feedback which is mostly coherent at low
temperatures and for moderate oscillator-environment coupling (in comparison to
the fundamental frequencies).

Our findings underline that non-Markovian effects are the key towards a deeper
understanding of this kind of entanglement dynamics. This is in
contrast to the behavior of subsystems coupled to independent heat baths, for
which non-Markovian effects are not essential, while thermal relaxation and
decoherence-free subspaces dominate. An interesting consequence of our results
in the realm of quantum information may be found in setups for quantum
communication and teleportation. Considering the studied model as a simplified
quantum network, our result for two-mode entanglement in the presence of
a passive oscillator imply the need for sufficient microscopic control of the
interaction between all constituents.  Thus, an interesting task would be the
prediction of the stability of such protocols under even weak interaction with
a common extended environment.

\acknowledgments
The authors warmly thank Luis A. Correa, Robert Hussein, Jos\'{e} P. Palao, and
Antonia Ruiz for fruitful discussions. A.A.V. would like to thank A. Castro
Castilla, and N. Garc\'{i}a Marco for many discussions on mathematical aspects.
This project was funded by the Spanish MICINN (Grant Nos.\ FIS2010-19998 and
MAT2011-24331) and by the European Union (FEDER). A.A.V. acknowledges financial
support by the Government of the Canary Islands through an ACIISI fellowship (85\%
co-financed by the European Social Fund).

\appendix

\section{The system-environment model}\label{App1}
In this appendix we derive the Langevin equation and different quantities used
in the main text. We start with the Hamiltonians $H_S$, $H_B$,
and $H_I$, Eqs.~\eqref{eq:1} and \eqref{eq:3}. We shall first neglect the
counter-term (renormalization) whose contribution will be included at the end.
Hence, the Hamiltonian equations of motion for $p_{\lambda} $ and $\ak$ are given by
\begin{eqnarray}
\dot p_{\lambda}&=&-m\omega_{\lambda}^{2}x_{\lambda}+\sum_{ k} \gk \left(\ak e^{i \vk \cdot \vrvo_{\lambda}}+\adk e^{-i \vk \cdot \vrvo_{\lambda}} \right)\label{Ham1},\\ 
\dot \ak&=&-i\wk  \ak {+}\frac{i}{\hbar} \sum_{\mu}\gk e^{-i \vk \cdot \vrvo_{\mu}}x_{\mu},
\label{Ham2}
\end{eqnarray}
where the latter possesses the formal solution
\begin{eqnarray}
\ak (t)&=& \ak (t_{0})e^{-i\wk (t-t_{0})}{+} \nonumber\\
&+&\frac{i}{\hbar} \sum_{\mu}\gk e^{-i \vk \cdot \vrvo_{\mu}}\int_{t_{0}}^{t}ds x_{\mu}(s)e^{-i\wk (t-s)}. \nonumber
\end{eqnarray}
We insert it into Eq.~\eqref{Ham1} to obtain for the oscillators
conditioned to the state of the environment the effective dynamical equation
\begin{equation}
\begin{split}
\dot p_{\lambda}
={}& -m\omega_{\lambda}^{2}x_{\lambda}+F_{\lambda}(t)
\\
& + \frac{i}{\hbar}
\sum_{\mu}\sum_{\vk} \gk^2 e^{ i \vk \cdot \left( \vrvo_{\lambda}-\vrvo_{\mu}  \right)} \int_{t_{0}}^{t}ds x_{\mu}(s)e^{-i \wk (t-s)}
\\
& -\frac{i}{\hbar}\sum_{\mu}\sum_{\vk}\gk^2 e^{-i \vk \cdot \left( \vrvo_{\lambda}-\vrvo_{\mu}  \right)}\int_{t_{0}}^{t}ds x_{\mu}(s)e^{i\wk (t-s)}.
\end{split}
\end{equation}
This equation can be expressed in a more convenient form by introducing the
fluctuating force $F_{\lambda}(t)$ and susceptibility $\chi_{\lambda \mu}(t)$ to
read
\begin{eqnarray}\nonumber
\dot p_{\lambda}(t)+m\omega_{\lambda}^{2}x_{\lambda}+\frac{1}{\hbar}{\displaystyle\int_{t_{0}}^{t}}d\tau \sum_{\mu}\chi_{\lambda \mu}(t-\tau)x_{\mu}(\tau)= F_{\lambda}(t),
\label{LGE1}
\end{eqnarray}
where
\begin{eqnarray}
\nonumber
F_{\lambda}(t)&=&
\sum_{\vk}\gk \Big(\ak (t_{0})e^{i( \vk \cdot \vrvo_{\lambda}-\wk (t-t_{0}))}
\\\nonumber && \qquad +
\ak ^{\dagger}(t_{0})e^{-i( \vk \cdot \vrvo_{\lambda}-\wk (t-t_{0}))}    \Big),\nonumber \\
\nonumber \\
\chi_{\lambda \mu}(t)&=& 2 \Theta\left(t-\Rr / c \right)\sum_{\vk}\gk^2  \sin\left(\vk \cdot \Rra - \wk  t \right). \nonumber
\end{eqnarray}
The susceptibility can be written in terms of an average over the environmental
state $\rho_B$ of the commutator of the fluctuating force, so that it becomes
\begin{equation}
\label{FDintime}
\chi_{\lambda \mu}\left( t-t'\right)
= - i\Theta( t-t' -\Rr /c)\big\langle\left[F_{\lambda}(t),F_{\mu}(t')
\right]\big\rangle_{\rho_{B}},
\end{equation}
where  $\Rr=|\vrvo_{\lambda}- \vrvo_{\mu}|$. 

The environment is initially in an equilibrium state at temperature $T$ for
which $\langle a^{\dagger}_{\vk'}\ak \rangle=\delta_{\vk \vk'}N(\wk )$, with the
bosonic thermal occupation $N(\wk) = [\exp(-\wk/k_BT)-1]^{-1}$ so that the
anti-commutator of the fluctuating force obeys
\begin{equation}
\begin{split}
\big\langle\left\lbrace F_{\lambda}(t),F_{\mu}(t') \right\rbrace \big\rangle_{\rho_{B}}
={}& 2 \sum_{\vk}\gk^2 {\Big( 2 N(\wk )+1 \Big)}
\\
& \times\cos\left[\vk \cdot \Rra-\wk (t-t') \right].
\end{split}
\end{equation}
In the frequency domain, this relation reads
\begin{equation}
\begin{split}
\big\langle\lbrace F_{\lambda}(\omega),F_{\mu}(\omega') \rbrace \big\rangle_{\rho_{B}}
={}&
4\pi^{2} \delta \left( \omega' + \omega \right) \coth\left( \frac{\hbar \omega'}{2k_{B}T}\right)
\\ & \times\sum_{\vk} \gk^2 
\Big( e^{ik \Delta x_{\lambda\mu}} \delta \left(\omega'-\wk  \right)
\\ & - e^{-ik
\Delta x_{\lambda\mu}} \delta\left(\omega'+\wk  \right) \Big) ,\label{FD1}
\end{split}
\end{equation}
where we have inserted $ 2 N(\wk )+1 =\coth\left( \hbar \omega/2K_{B} T\right)$.
For a more compact notation, we introduce the spectral densities
\begin{equation}
J_{\lambda{,}\mu}(\omega)= \frac{\pi}{\hbar}\sum_{\vk} \gk^2 \cos\left( \vk \cdot \Rra \right) \delta\left(\omega-\wk  \right),\label{SpecDens}
\end{equation}
with which we obtain from Eq.~\eqref{FD1} the quantum fluctuation-dissipation
relation
\begin{equation}
\Re\frac{1}{2}{ \Big\langle\left\lbrace{ F_{\lambda}(\omega),F_{\mu}(\omega') }\right\rbrace \Big\rangle }= 2 \pi \hbar \delta \left(\omega' + \omega \right)\Gamma_{\lambda\mu}(\omega'),
\end{equation}
with the imaginary part of the susceptibility
\begin{align}
\Gamma_{\lambda\mu}(\omega)
={}& -\frac{1}{\hbar}\hbox{Im }\chi_{\lambda\mu}(\omega){ \coth\left( \frac{\hbar
\omega'}{2 K_{B} T}\right)} \label{ImJ}\\
={}&
J_{\lambda,\mu}(|\omega|)\,
{\coth\left( \frac{\hbar|\omega|}{2 K_{B} T}\right)},
\nonumber
\end{align}
derived in Appendix B.

So far we have not taken into account the counter-term. In doing so, the spectral
densities lead to harmonic renormalization potentials with frequencies
\begin{eqnarray}
\tilde{\Omega}_{\lambda \lambda}&=&\frac{1}{\hbar}\sum_{\vk} \frac{\gk^{2}}{\wk }=\frac{1}{\pi }\int_{0}^{\infty}\frac{J_{\lambda,\lambda}(\omega)}{\omega}d\omega, \label{app:Omega} \\
\tilde{\Omega}_{\lambda\mu}&=&\frac{1}{\hbar}\sum_{\vk} \frac{\gk^{2} }{\wk }\cos\left( \vk \cdot \Rra \right) =\frac{1}{\pi }\int_{0}^{\infty}\frac{J_{\lambda,\mu}(\omega)}{\omega}d\omega .
\nonumber
\end{eqnarray}
Owing to the linearity of the dynamical equations for $x_{\lambda}$ and
$p_{\lambda}$, it is straightforward to show that including the
counter-term provides the Langevin equation (\ref{GLE}).

\section{Spectral densities and susceptibilities}\label{App2}

Irrespective of the dimension of the environment, we assume that it is isotropic
and possesses the linear dispersion relation $\wk=c |\vk|$ with cut-off
frequency $\wc$. We model this by introducing coupling constants $g_{\vk}$ that
obey
\begin{equation}
g^2_{\vk}=m \hbar \gamma
(\wk/\omega_c^{d-1})c^{d} V_{\vk}(d) e^{-\omega/\omega_c},
\label{app:b1}
\end{equation}
where $d$ is the dimension of the environment, $V_{\vk}$ is the $d$-dimensional
$\vk$-space volume per field mode, and $\gamma$ is the effective coupling
strength. We start from Eq.~\eqref{SpecDens} and take the continuum limit
$V_{\vk}\to 0$. We provide explicit expressions for the dimensions $d=1$ and
$d=3$, while $d=2$ is addressed mainly for highlighting the difficulties that
arise in that dimension.

\subsection{One-dimensional environment}

Inserting Eq.~\eqref{app:b1} for $d=1$ into \eqref{SpecDens} and
\eqref{app:Omega} yields in the continuum limit
$V_{\vk}(1) \rightarrow 0$ for the spectral density the closed-form form
\begin{equation}
J_{\lambda\mu}(\omega)=\pi  m \gamma \omega e^{-\omega/\wc}\cos ( \omega \Rr/c ), 
\end{equation}
and the potential renormalization frequencies
\begin{eqnarray}\nonumber
\tilde{\Omega}_{\lambda \lambda}&=& m \gamma \omega_{c}, \\
\tilde{\Omega}_{\lambda \mu}&=&  \frac{m \gamma \omega_{c}}{ 1+(\wc \Rr/c)^{2} },\nonumber
\end{eqnarray}
respectively.
The real part of the susceptibility ${\chi_{\lambda \mu}(\omega')}$ is
obtained from Eq.~\eqref{ImJ} via the Kramers-Kronig relations.  Mathematically
this corresponds to the Hilbert transformation \cite{erdelyi1954}can formally be
expressed as
\begin{equation}
\begin{split}
\Re{\chi_{\lambda \mu}(\omega')}
={}& \mathcal{H}\left[\Im\chi_{\lambda\mu}(\omega)\right] \left( \omega'\right)
\\
:={}& \frac{1}{\pi}P\int_{-\infty}^{\infty}\frac{\Im\chi_{\lambda
\mu}(\omega)}{\omega-\omega'}d\omega ,
\end{split}
\end{equation}
where $P$ is the Cauchy principal value and $\mathcal{H}[f(\omega)](\omega')$
the Hilbert transform of $f(\omega)$. Hence,
\begin{equation}
\begin{split}
\Re{\chi_{\lambda \mu}(\omega)}
={}& -m  \hbar \gamma P \int_{0}^{\infty}\omega e^{-\omega/\wc}\cos(\omega \Rr/c)
\\&\times
\Big( \frac{1}{\omega-\omega'}+\frac{1}{\omega+\omega'} \Big)d\omega,
\end{split}
\label{Hilbert1}
\end{equation} 
which consists of two terms that differ by the sign of $\omega'$ and, thus, it
is sufficient to compute
\begin{widetext}
\begin{eqnarray}\nonumber
P\int_{0}^{\infty}\frac{\omega e^{-\omega/\wc}\cos( \omega \Rr/c)}{\omega-\omega'}
d\omega
= \omega'P\int_{0}^{\infty}\frac{ e^{-\omega/\wc}\cos(\omega \Rr/c)
}{\omega-\omega'}d\omega+\frac{\omega_{c}}{1+(\wc \Rr/c)^{2}},
\end{eqnarray}
where we have used $\mathcal{H}[\omega f(\omega)]=\omega
\mathcal{H}(f(\omega))+\frac{1}{\pi}\int_{-\infty}^{\infty}f(\omega)d\omega$ to
arrive at
\begin{eqnarray}
P\int_{0}^{\infty}\frac{ e^{-\left(\frac{1}{\omega_{c}}-i \Rr/c \right)\omega }}{\omega-\omega'}d\omega =
\begin{cases}
{e^{-\left(\frac{1}{\omega_{c}}-i \Rr/c \right)\omega' }\left\{  \Gamma\left[0,-
\left(\frac{1}{\omega_{c}}-i\Rr/c \right)\omega' \right] +i \pi\right\} }& \mbox{if } \, \omega'\, \in (0,\infty)    \\
{e^{-\left(\frac{1}{\omega_{c}}-i\Rr/c \right)\omega' } \Gamma\left[0,- \left(\frac{1}{\omega_{c}}-i\Rr/c \right)\omega' \right]}& \mbox{if } \, \omega'\, \in (-\infty,0)    .\nonumber
\end{cases}
\end{eqnarray}
where $\Gamma(a,z)=\int^{z}_{\infty}t^{a-1}e^{-t}dt$ denotes the incomplete
gamma function.
Inserting this expression into Eq.~\eqref{Hilbert1}, we finally obtain
\begin{eqnarray}
\Re{\chi_{\lambda \mu}(\omega)}
&=&-m \hbar \gamma \omega \Re\left[ g(\omega)-g(-\omega)\right]
+ \pi m \hbar \, \gamma \, \omega \, \Im\big[\Theta\left(\omega \right)
e^{-\left(\frac{1}{\omega_{c}}-i \Rr/c \right)\omega}
\nonumber\\&&
-\Theta\left(-\omega
\big] e^{\left(\frac{1}{\omega_{c}}-i \Rr/c \right)\omega} \right)
-\frac{2 m\hbar\gamma \omega_{c} }{1+(\wc \Rr /c)^{2} } ,
\end{eqnarray}
with $g(\omega)= e^{-(1-i \wc \Rr /c)\omega /\wc} \Gamma[0,-(1-i \wc \Rr
/c)\omega /\wc]$. From this expression we find the well known relation
between the frequency shift $\Delta \omega_{\lambda \mu} $ and the real part of
susceptibility \cite{riseborough19851},
\begin{equation}\nonumber
\left( \Delta \omega_{\lambda \mu}\right)^2=-\frac{\tilde{\Omega}_{\lambda
\mu}}{m}=\frac{1}{2m \hbar}\lim_{\omega\rightarrow 0}\Re{\chi_{\lambda \mu}(\omega)}.
\end{equation}
\end{widetext}

\subsection{Two-dimensional environment}

Again, we use \eqref{app:b1}, perform the continuum limit, and readily obtain
\begin{eqnarray}
J_{\lambda\mu}(\omega)=2 \pi^{2}m \gamma  \frac{\omega^{2}}{\omega_{c}}
e^{-\omega/\omega_{c}}\mathsf{J}_{0}( \omega \Rr /c) ,
\end{eqnarray}
where $\mathsf{J}_{0}$ is the zeroth-order Bessel function of the first kind. The
renormalization frequencies now become
\begin{eqnarray}\nonumber
\tilde{\Omega}_{\lambda \lambda}&=& 2 \pi m \gamma \omega_{c} ,\\
\tilde{\Omega}_{\lambda \lambda'}&=& \frac{2\pi m \gamma \omega_{c}}{[
1+( \wc \Rr/c)^{2}]^{3/2} } . \nonumber
\end{eqnarray}
Accordingly, the Fourier transform of the real part of the susceptibility reads
\begin{equation}
\begin{split}
\Re\chi_{\lambda \mu}(\omega')
={}& -\frac{2 \pi m \gamma \hbar}{\omega_{c}} P\int_{0}^{\infty}
\omega^{2} e^{-\omega/\omega_{c}}
\mathsf{J}_{0}(\omega \Rr /c) 
\\ &\times
\left(\frac{1}{\omega-\omega'}+\frac{1}{\omega+\omega'}\right)d\omega.
\end{split}
\end{equation}
Using the same relation of Hilbert transforms as in the previous
section we can write,
\begin{equation}
\begin{split}
\mathcal{H}[ & \Theta(\omega)\omega^{2} e^{-\omega/\omega_{c}}\mathsf{J}_{0}(\omega
\Rr /c) ] (\omega' )
\\
={} & \omega'^{2}\mathcal{H}[\Theta(\omega) e^{-\omega/\omega_{c}}\mathsf{J}_{0}(
\omega \Rr /c) ] (\omega') \nonumber\\
&+\frac{\omega'\omega_{c}}{[ 1+( \wc \Rr /c) ^{2}]^{1/2}}
+\frac{\omega_{c}^{2}}{[ 1+( \wc \Rr /c) ^{2}]^{3/2}}.
\end{split}
\end{equation}
Here a major difficulty arises.  The Hilbert transform 
$\mathcal{H}[\Theta(\omega) e^{-\omega/\omega_{c}}\mathsf{J}_{0}({\omega
\Rr}/{c})](\omega')$ exists only for $R/c=1$, despite the convergence condition
$0<\omega_{c}$.  Thus, we cannot derive any closed expression for
$\Re\chi(\omega)$ for all $R$ and $c$. Still we obtain
by using a series representation for
$\mathsf{J}_{0}({\omega \Rr}/{c})$ the relation
\begin{equation}
\begin{split}
\Re{\chi(\omega)}
={} &\frac{2 \pi m \gamma \hbar}{\omega_{c}}\omega^ {2}
\Big[\Theta(\omega)e^{\frac{-\omega}{\omega_{c}}}Ei\left(\frac{\omega}{\omega_{c}}
\right)
\\
& -\Theta(-\omega)e^{\frac{\omega}{\omega_{c}}}Ei\left(\frac{-\omega}{\omega_{c}}
\right) \Big]\mathsf{J}_{0}({{\Rr} \omega}/{c} )
\\
& -2\pi m \gamma \hbar \omega_{c}\sum_{l=0}^{\infty}\frac{(-1)^l}{2^{2l}}\left(
\frac{{\Rr} \omega_{c}}{c}\right)^{2l}
\\ & \times \sum_{k=1}^{l+1}
\frac{(2(l-k)+3)!}{l!l!}\left( \frac{\omega}{\omega_{c}}\right)^{2(k-1)}.
\end{split}
\end{equation}
This series, however, it is not of practical use, because of its slow
convergence.

\subsection{Three-dimensional environment}

Following once more the same line, we obtain the spectral densities
\begin{equation}
J_{\lambda\mu}(\omega)=4\pi^{2} m  \gamma \frac{c}{\Rr}\Big(\frac{\omega}{\omega_{c}}\Big)^{2}\sin(\omega \Rr /c )e^{-\omega/\wc},
\end{equation}
and the renormalization frequencies
\begin{eqnarray}\nonumber
\tilde{\Omega}_{\lambda \lambda}&=& 8 \pi  m \gamma \omega_{c} \nonumber \\
\tilde{\Omega}_{\lambda \mu}&=& \frac{8 \pi m  \gamma \omega_{c}}{[ 1+( \wc \Rr
/c )^{2}]^2}. \nonumber
\end{eqnarray}
Now the real part of the susceptibility is given by
\begin{equation}
\begin{split}
\Re{\chi_{\lambda \mu}(\omega')}
={} & -\frac{4\pi m \hbar  \gamma}{\omega_{c}^{2}}\Big(\frac{c}{\Rr}\Big) P\int_{0}^{\infty}
\sin(\omega \Rr/ c)
\\ & \times
\omega^{2}e^{-\omega/\omega_{c}}\Big(
\frac{1}{\omega-\omega'}+ \frac{1}{\omega+\omega'}\Big) d\omega,
\end{split}
\end{equation}
where the integral can be written as
\begin{equation}
\begin{split}
{\frac{1}{\pi}} P & \int_{0}^{\infty}\frac{\omega^{2} \sin(\omega \Rr/c)e^{\frac{-\omega}{\omega_{c}}}}{\omega-\omega'}d\omega
\\ ={} &
\omega'^{2}\mathcal{H}\left[ \Theta(\omega)\sin(\omega \Rr/c)e^{\frac{-\omega}{\omega_{c}}}\right] ( \omega')
\\ &
+{\frac{\omega'}{\pi}}\frac{\wc^{2} \Rr /c}{[1+ ( \wc \Rr/c)^{2} ]}
\\ & +{\frac{1}{\pi}}\frac{ 2\omega_{c}^{3} \Rr/c}{[ 1+( \wc \Rr/c )^{2}]^{2} } .
\end{split}
\end{equation}
After some algebra, we finally obtain for the real part of the 3D-susceptibility
the expression
\begin{equation}
\begin{split}
\Re{\chi_{\lambda \nu}(\omega)}
={} &{-\frac{4\pi m \hbar \gamma c \omega^{2}}{\omega_{c}^{2}{\Rr}}
    \Im\left[g(\omega)+ g(-\omega)\right]}  \\
&- \frac{4\pi^2 m \hbar\gamma c \omega^{2}}{\omega_{c}^{2}{\Rr}}
   \Re \big[\Theta(\omega) e^{-(\frac{1}{\omega_{c}}-i \Rr/c) \omega}
\\ & 
+\Theta(-\omega) e^{(\frac{1}{\omega_{c}}-i \Rr/c )\omega} \big]
\\
&-\frac{16\pi m  \hbar \gamma \omega_{c} }{[ 1+( \wc\Rr /c)^{2}]^{2} }
\end{split}
\end{equation}
with $g(\omega)= e^{-(1-i \wc \Rr /c)\omega /\wc} \Gamma[0,-(1-i \wc \Rr
/c)\omega /\wc]$ and the incomplete gamma function $\Gamma(0,x)$.

\section{Fourier representation of Eq.~\eqref{FD1}}

Here a give a simple proof of Eq.~\eqref{ImJ} starting from the Fourier transform
of the susceptibility
\begin{eqnarray}\nonumber
\chi_{\lambda \mu}\left( \omega\right)
&=&\int_{-\infty}^{\infty} e^{i\omega t}\chi_{\lambda \mu}(t) \,dt
\\
&=& 2 \int\limits_{ \Rr /c}^{\infty} e^{i\omega t}\sum_{\vk}\gk^{2}\sin\left( \vk\cdot \Rra-\wk t\right) \,dt \nonumber\\
&=& -i \sum_{\vk}\gk^{2}\Bigg[e^{i \left( \vk\cdot\Rra- \left( \omega-\wk
\right)\Rr/c \right)}
\nonumber\\ && \times
\int\limits_{0}^{\infty}e^{i\left(\omega-\wk \right)t}\,dt- e^{-i
\left( \vk\cdot\Rra- \left( \omega+\wk \right)\Rr/c \right)}
\nonumber\\ && \times
\int\limits_{0}^{\infty}e^{i\left(\omega+\wk \right)t} \,dt \Bigg] ,\nonumber
\label{ft}
\end{eqnarray}
where we have made the substitution $t\to t+\Rr/c$. Inserting
\begin{equation}\nonumber
\int_{0}^{\infty}e^{i\left(\omega-\wk  \right) t}\,dt =\pi\delta\left( \omega-\wk \right) +i\mathcal{H}\left( 1 \right) \left( \wk  \right) 
\end{equation}
into Eq.~\eqref{ft} yields
\begin{equation}
\nonumber
\begin{split}
\chi_{\lambda \mu}\left( \omega\right)
={} &-i\pi \sum_{\vk}\gk^{2}\Big\{
e^{i \left( \vk\cdot \Rra-\left( \omega-\wk \right)\Rr/c \right)}
\\&\times
\delta( \omega-\wk) -e^{-i (\vk\cdot \Rra-( \omega+\wk)\Rr/c)}
\\&\times
\delta( \omega+\wk) \Big\}
\\&
+\sum_{\vk}\gk^{2}\Big\{e^{i \left( \vk\cdot \Rra-\left( \omega-\wk \right)\Rr/c\right)}
\mathcal{H}(1) \left( \wk  \right)
\\&
- e^{-i \left( \vk\cdot \Rra-\left( \omega+\wk \right)\Rr/c \right)}
\mathcal{H}\left( 1\right)\left( \wk  \right) \Big\}, \nonumber
\end{split}
\end{equation}
where the second sum vanishes owing to $\mathcal{H}(1) ( \wk )=0$
\cite{erdelyi1954}.  By taking the imaginary part and performing the continuum
limit, we obtain Eq.~\eqref{ImJ}.

\section{PPT criterion and classification of tripartite entanglement}\label{sectionppt}

Let us consider a system composed of two parties $ A$ and $ B$. Then a necessary
and sufficient condition for the separability between $1_{A} \times 1_{B}$ (two
modes), $1_{A} \times N_{ B}$, and $N_{A}\times N_{B}$ bisymmetric bipartite
states is the partial positive transpose (PPT) criterion \cite{adesso20071,
serafini20051}. The $N_{A} \times N_{B}$ class of systems relates to Gaussian
states that are locally invariant under all permutations of modes in each of the
two subsystems. Then the PPT criterion can be formulated in terms of a
bisymmetric covariance matrix $\ten G$ as follows: A state is separable if and
only if $\ten G^ {T_{B}} \geq ({i} \hbar/{2}) \ten\sigma$ (i.e., $\ten G^
{T_{B}} $ is a positive-definite matrix), where $\ten G^ {T_{ B}}$ is the
covariance matrix of the partial transpose of $\ten G$ with respect to the
system $ B$, given by
$ 
\ten G^ {T_{B}}=:\ten \Lambda \ten G \ten\Lambda,
$ 
with 
\begin{align}\nonumber
\ten\Lambda=\mathbb{I}_{N_{ A}+N_{B}}\oplus 
\begin{bmatrix}
     \mathbb{I}_{N_{ A}} &  0\\
      0 & -\mathbb{I}_{N_{ B}} \\
\end{bmatrix} ,
\end{align}
the $N$-dimensional unit matrix $\mathbb{I}_N$,
and the symplectic matrix
\begin{align}\nonumber
\ten\sigma=\begin{bmatrix}
     0 &  \mathbb{I}_{N_{ A}+N_{ B}}\\
      -\mathbb{I}_{N_{ A}+N_{ B}}.& 0 \\
\end{bmatrix}\,.
\end{align}
The PPT criterion can readily be evaluated from the symplectic eigenvalues of
$\ten G^ {T_{ B}} $, given by the positive square roots of the eigenvalues of
$(-i/\hbar) \ten \sigma \ten G^{T_{B}} $ \cite{vidal20021}.

For a system composed of three-modes, Giedke \textit{et al.}\ \cite{giedke20011} have
considered the PPT criterion to provide a complete classification of the
three-mode states, according their separability properties. This classification
is based on the partially transposed covariance matrices $\tilde{\ten G}^
{T_{\lambda}}=\ten\Lambda_{\lambda} \ten G \ten\Lambda_{\lambda}$, which is
related to the three possible bi-partitions of a three-component system, namely
$\A|\B\C$, $\A \B |\C$ and $\A\C|\B$. Then each three-mode Gaussian state can
be assigned to one of the following classes \cite{giedke20011}:
\begin{enumerate}
\item[C1]\textit{Fully inseparable states} that are not separable
under any of the three possible bipartitions.  This class contains the
\textit{genuine tripartite entangled} states \cite{benatti20111}.

\item[C2]\textit{One-mode biseparable states} that are separable if two of the
parties are grouped together, but inseparable with respect to the other
groupings. 

\item[C3]\textit{Two-mode biseparable states} for which two of the bipartitions
are separable.

\item[C4]\textit{Three-mode biseparable states} for which all the three
bipartitions are separable, but which cannot be written as a
mixture of tripartite product states. These states are also known as
\textit{tripartite bound entangled states}.

\item[C5]\textit{Fully separable states} that can be written as a mixture of
tripartite product states.
\end{enumerate}


\begin{thebibliography}{10}%
\makeatletter
\providecommand \@ifxundefined [1]{%
 \ifx #1\undefined \expandafter \@firstoftwo
 \else \expandafter \@secondoftwo
\fi
}%
\providecommand \@ifnum [1]{%
 \ifnum #1\expandafter \@firstoftwo
 \else \expandafter \@secondoftwo
\fi
}%
\providecommand \enquote [1]{``#1''}%
\providecommand \bibnamefont  [1]{#1}%
\providecommand \bibfnamefont [1]{#1}%
\providecommand \citenamefont [1]{#1}%
\providecommand\href[0]{\@sanitize\@href}%
\providecommand\@href[1]{\endgroup\@@startlink{#1}\endgroup\@@href}%
\providecommand\@@href[1]{#1\@@endlink}%
\providecommand \@sanitize [0]{\begingroup\catcode`\&12\catcode`\#12\relax}%
\@ifxundefined \pdfoutput {\@firstoftwo}{%
 \@ifnum{\z@=\pdfoutput}{\@firstoftwo}{\@secondoftwo}%
}{%
 \providecommand\@@startlink[1]{\leavevmode}%
 \providecommand\@@endlink[0]{}%
}{%
 \providecommand\@@startlink[1]{%
  \leavevmode
  \pdfstartlink
   attr{/Border[0 0 1 ]/H/I/C[0 1 1]}%
   user{/Subtype/Link/A<</Type/Action/S/URI/URI(#1)>>}%
  \relax
 }%
 \providecommand\@@endlink[0]{\pdfendlink}%
}%
\providecommand \url  [0]{\begingroup\@sanitize \@url }%
\providecommand \@url [1]{\endgroup\@href {#1}{\urlprefix}}%
\providecommand \urlprefix [0]{URL }%
\providecommand \Eprint[0]{\href }%
\@ifxundefined \urlstyle {%
  \providecommand \doi [1]{doi:\discretionary{}{}{}#1}%
}{%
  \providecommand \doi [0]{doi:\discretionary{}{}{}\begingroup
  \urlstyle{rm}\Url }%
}%
\providecommand \doibase [0]{http://dx.doi.org/}%
\providecommand \Doi[1]{\href{\doibase#1}}%
\providecommand \bibAnnote [3]{%
  \BibitemShut{#1}%
  \begin{quotation}\noindent
    \textsc{Key:}\ #2\\\textsc{Annotation:}\ #3%
  \end{quotation}%
}%
\providecommand \bibAnnoteFile [2]{%
  \IfFileExists{#2}{\bibAnnote {#1} {#2} {\input{#2}}}{}%
}%
\providecommand \typeout [0]{\immediate \write \m@ne }%
\providecommand \selectlanguage [0]{\@gobble}%
\providecommand \bibinfo [0]{\@secondoftwo}%
\providecommand \bibfield [0]{\@secondoftwo}%
\providecommand \translation [1]{[#1]}%
\providecommand \BibitemOpen[0]{}%
\providecommand \bibitemStop [0]{}%
\providecommand \bibitemNoStop [0]{.\EOS\space}%
\providecommand \EOS [0]{\spacefactor3000\relax}%
\providecommand \BibitemShut [1]{\csname bibitem#1\endcsname}%
\bibitem{horodecki20091}%
  \BibitemOpen
  \bibfield{author}{%
  \bibinfo {author} {\bibfnamefont{R.}~\bibnamefont{Horodecki}}, \bibinfo
  {author} {\bibfnamefont{P.}~\bibnamefont{Horodecki}}, \bibinfo {author}
  {\bibfnamefont{M.}~\bibnamefont{Horodecki}},\ and\ \bibinfo {author}
  {\bibfnamefont{K.}~\bibnamefont{Horodecki}},\ }%
  \bibfield{journal}{%
  \Doi{10.1103/RevModPhys.81.865}{\bibinfo {journal} {Rev. Mod. Phys.}}\ }%
  \textbf{\bibinfo {volume} {81}},\ \bibinfo {pages} {865} (\bibinfo {year} {2009}).
\bibitem{zurek2003}%
  W. H. Zurek, Rev. Mod. Phys. 75, 715 (2003).
\bibitem{braun20021}%
  \BibitemOpen
  \bibfield{author}{%
  \bibinfo {author} {\bibfnamefont{D.}~\bibnamefont{Braun}},\ }%
  \bibfield{journal}{%
  \Doi{10.1103/PhysRevLett.89.277901}{\bibinfo {journal} {Phys. Rev. Lett.}}\
  }%
  \textbf{\bibinfo {volume} {89}},\ \bibinfo {pages} {277901} (\bibinfo {year} {2002}).
\bibitem{plenio20021}%
  \BibitemOpen
  \bibfield{author}{%
  \bibinfo {author} {\bibfnamefont{M.~B.}\ \bibnamefont{Plenio}}\ and\ \bibinfo
  {author} {\bibfnamefont{S.~F.}\ \bibnamefont{Huelga}},\ }%
  \bibfield{journal}{%
  \Doi{10.1103/PhysRevLett.88.197901}{\bibinfo {journal} {Phys. Rev. Lett.}}\
  }%
  \textbf{\bibinfo {volume} {88}},\ \bibinfo {pages} {197901} (\bibinfo {year} {2002})

\bibitem{benatti20031}%
  \BibitemOpen
  \bibfield{author}{%
  \bibinfo {author} {\bibfnamefont{F.}~\bibnamefont{Benatti}}, \bibinfo
  {author} {\bibfnamefont{R.}~\bibnamefont{Floreanini}},\ and\ \bibinfo
  {author} {\bibfnamefont{M.}~\bibnamefont{Piani}},\ }%
  \bibfield{journal}{%
  \Doi{10.1103/PhysRevLett.91.070402}{\bibinfo {journal} {Phys. Rev. Lett.}}\
  }%
  \textbf{\bibinfo {volume} {91}},\ \bibinfo {pages} {070402} (\bibinfo {year} {2003}).
  
\bibitem{benatti20101}%
  \BibitemOpen
  \bibfield{author}{%
  \bibinfo {author} {\bibfnamefont{F.}~\bibnamefont{Benatti}}, \bibinfo
  {author} {\bibfnamefont{R.}~\bibnamefont{Floreanini}},\ and\ \bibinfo
  {author} {\bibfnamefont{U.}~\bibnamefont{Marzolino}},\ }%
  \bibfield{journal}{%
  \Doi{10.1103/PhysRevA.81.012105}{\bibinfo {journal} {Phys. Rev. A}}\
  }%
  \textbf{\bibinfo {volume} {81}},\ \bibinfo {pages} {012105} (\bibinfo {year} {2010}).
  
\bibitem{doll20061}%
  \BibitemOpen
  \bibfield{author}{%
  \bibinfo {author} {\bibfnamefont{R.}~\bibnamefont{Doll}}, \bibinfo {author}
  {\bibfnamefont{M.}~\bibnamefont{Wubs}}, \bibinfo {author}
  {\bibfnamefont{P.}~\bibnamefont{Hänggi}},\ and\ \bibinfo {author}
  {\bibfnamefont{S.}~\bibnamefont{Kohler}},\ }%
  \bibfield{journal}{%
  \bibinfo {journal} {EPL (Europhysics letter)}\ }%
  \textbf{\bibinfo {volume} {76}},\ \bibinfo {pages} {547} (\bibinfo {year}
  {2006}).
\bibitem{shiokawa20091}%
  \BibitemOpen
  \bibfield{author}{%
  \bibinfo {author} {\bibfnamefont{K.}~\bibnamefont{Shiokawa}},\ }%
  \bibfield{journal}{%
  \Doi{10.1103/PhysRevA.79.012308}{\bibinfo {journal} {Phys. Rev. A}}\ }%
  \textbf{\bibinfo {volume} {79}},\ \bibinfo {pages} {012308} (\bibinfo {year} {2009}).

\bibitem{wolf20111}%
  \BibitemOpen
  \bibfield{author}{%
  \bibinfo {author} {\bibfnamefont{A.}~\bibnamefont{Wolf}}, \bibinfo {author}
  {\bibfnamefont{G.~D.}\ \bibnamefont{Chiara}}, \bibinfo {author}
  {\bibfnamefont{E.}~\bibnamefont{Kajari}}, \bibinfo {author}
  {\bibfnamefont{E.}~\bibnamefont{Lutz}},\ and\ \bibinfo {author}
  {\bibfnamefont{G.}~\bibnamefont{Morigi}},\ }%
  \bibfield{journal}{%
  \bibinfo {journal} {EPL (Europhysics letter)}\ }%
  \textbf{\bibinfo {volume} {95}},\ \bibinfo {pages} {60008} (\bibinfo {year}
  {2011}).
\bibitem{kajari20121}%
  \BibitemOpen
  \bibfield{author}{%
  \bibinfo {author} {\bibfnamefont{E.}~\bibnamefont{Kajari}}, \bibinfo {author}
  {\bibfnamefont{A.}~\bibnamefont{Wolf}}, \bibinfo {author}
  {\bibfnamefont{E.}~\bibnamefont{Lutz}},\ and\ \bibinfo {author}
  {\bibfnamefont{G.}~\bibnamefont{Morigi}},\ }%
  \bibfield{journal}{%
  \Doi{10.1103/PhysRevA.85.042318}{\bibinfo {journal} {Phys. Rev. A}}\ }%
  \textbf{\bibinfo {volume} {85}},\ \bibinfo {pages} {042318} (\bibinfo {year} {2012}).
  \bibitem{doll20071}%
  \BibitemOpen
  \bibfield{author}{%
  \bibinfo {author} {\bibfnamefont{R.}~\bibnamefont{Doll}}, \bibinfo {author}
  {\bibfnamefont{M.}~\bibnamefont{Wubs}}, \bibinfo {author}
  {\bibfnamefont{P.}~\bibnamefont{H\"anggi}},\ and\ \bibinfo {author}
  {\bibfnamefont{S.}~\bibnamefont{Kohler}},\ }%
  \bibfield{journal}{%
  \Doi{10.1103/PhysRevB.76.045317}{\bibinfo {journal} {Phys. Rev. B}}\ }%
  \textbf{\bibinfo {volume} {76}},\ \bibinfo {pages} {045317} (\bibinfo {year} {2007}).
\bibitem{horhammer20081}%
  \BibitemOpen
  \bibfield{author}{%
  \bibinfo {author} {\bibfnamefont{C.}~\bibnamefont{H\"orhammer}}\ and\
  \bibinfo {author} {\bibfnamefont{H.}~\bibnamefont{B\"uttner}},\ }%
  \bibfield{journal}{%
  \Doi{10.1103/PhysRevA.77.042305}{\bibinfo {journal} {Phys. Rev. A}}\ }%
  \textbf{\bibinfo {volume} {77}},\ \bibinfo {pages} {042305} (\bibinfo {year} {2008}).
\bibitem{zell20091}%
  \BibitemOpen
  \bibfield{author}{%
  \bibinfo {author} {\bibfnamefont{T.}~\bibnamefont{Zell}}, \bibinfo {author}
  {\bibfnamefont{F.}~\bibnamefont{Queisser}},\ and\ \bibinfo {author}
  {\bibfnamefont{R.}~\bibnamefont{Klesse}},\ }%
  \bibfield{journal}{%
  \Doi{10.1103/PhysRevLett.102.160501}{\bibinfo {journal} {Phys. Rev. Lett.}}\
  }%
  \textbf{\bibinfo {volume} {102}},\ \bibinfo {pages} {160501} (\bibinfo {year} {2009}).
	
 
\bibitem{vasile20101}%
  \BibitemOpen
  \bibfield{author}{%
  \bibinfo {author} {\bibfnamefont{Ruggero}~\bibnamefont{Vasile}}, \bibinfo {author}
  {\bibfnamefont{Paolo}~\bibnamefont{Giorda}}, \bibinfo {author}
  {\bibfnamefont{Stefano}~\bibnamefont{Olivares}},\ \bibinfo {author}
  {\bibfnamefont{Matteo G.A.}\ \bibnamefont{Paris}}, \ and \ \bibinfo {author}{\bibfnamefont{Sabrina}\ \bibnamefont{Maniscalco}},\ }%
  \bibfield{journal}{%
  \Doi{10.1103/PhysRevA.82.012313}{\bibinfo {journal} {Phys. Rev. A}}\ }%
  \textbf{\bibinfo {volume} {82}},\ \bibinfo {pages} {012313} (\bibinfo {year} {2010}).	
	
	
	 \bibitem{fleming20121}%
  \BibitemOpen
  \bibfield{author}{%
  \bibinfo {author} {\bibfnamefont{C. H.}~\bibnamefont{Fleming}}, \bibinfo {author}{\bibfnamefont{N. I.}~\bibnamefont{Cummings}},  \bibinfo {author}{\bibfnamefont{Charis I.}~\bibnamefont{Anastopoulos}} and\ \bibinfo {author} {\bibfnamefont{B. L.}~\bibnamefont{Hu}},\ }%
  \bibfield{journal}{%
  \bibinfo {journal} {J. Phy. A: Math. and Theor.}\ }%
  \textbf{\bibinfo {volume} {45}},\ \bibinfo {pages} {065301} (\bibinfo {year}
  {2012}).
	
	
\bibitem{correa20121}%
  \BibitemOpen
  \bibfield{author}{%
  \bibinfo {author} {\bibfnamefont{L.~A.}\ \bibnamefont{Correa}}, \bibinfo
  {author} {\bibfnamefont{A.~A.}\ \bibnamefont{Valido}},\ and\ \bibinfo
  {author} {\bibfnamefont{D.}~\bibnamefont{Alonso}},\ }%
  \bibfield{journal}{%
  \Doi{10.1103/PhysRevA.86.012110}{\bibinfo {journal} {Phys. Rev. A}}\ }%
  \textbf{\bibinfo {volume} {86}},\ \bibinfo {pages} {012110} (\bibinfo {year} {2012}).
  \bibitem{duarte20091}%
  \BibitemOpen
  \bibfield{author}{%
  \bibinfo {author} {\bibfnamefont{O.~S.}\ \bibnamefont{Duarte}}\ and\ \bibinfo
  {author} {\bibfnamefont{A.~O.}\ \bibnamefont{Caldeira}},\ }%
  \bibfield{journal}{%
  \Doi{10.1103/PhysRevA.80.032110}{\bibinfo {journal} {Phys. Rev. A}}\ }%
  \textbf{\bibinfo {volume} {80}},\ \bibinfo {pages} {032110} (\bibinfo {year} {2009}).
\bibitem{valente20101}%
  \BibitemOpen
  \bibfield{author}{%
  \bibinfo {author} {\bibfnamefont{D.~M.}\ \bibnamefont{Valente}}\ and\
  \bibinfo {author} {\bibfnamefont{A.~O.}\ \bibnamefont{Caldeira}},\ }%
  \bibfield{journal}{%
  \Doi{10.1103/PhysRevA.81.012117}{\bibinfo {journal} {Phys. Rev. A}}\ }%
  \textbf{\bibinfo {volume} {81}},\ \bibinfo {pages} {012117} (\bibinfo {year} {2010}).

\bibitem{krauter20111}%
  \BibitemOpen
  \bibfield{author}{%
  \bibinfo {author} {\bibfnamefont{H.}~\bibnamefont{Krauter}}, \bibinfo
  {author} {\bibfnamefont{C.~A.}\ \bibnamefont{Muschik}}, \bibinfo {author}
  {\bibfnamefont{K.}~\bibnamefont{Jensen}}, \bibinfo {author}
  {\bibfnamefont{W.}~\bibnamefont{Wasilewski}}, \bibinfo {author}
  {\bibfnamefont{J.~M.}\ \bibnamefont{Petersen}}, \bibinfo {author}
  {\bibfnamefont{J.~I.}\ \bibnamefont{Cirac}},\ and\ \bibinfo {author}
  {\bibfnamefont{E.~S.}\ \bibnamefont{Polzik}},\ }%
  \bibfield{journal}{%
  \Doi{10.1103/PhysRevLett.107.080503}{\bibinfo {journal} {Phys. Rev. Lett.}}\ }%
  \textbf{\bibinfo {volume} {107}},\ \bibinfo {pages} {080503} (\bibinfo {year} {2011}).

\bibitem{leggett19871}%
  \BibitemOpen
  \bibfield{author}{%
  \bibinfo {author} {\bibfnamefont{A.~J.}\ \bibnamefont{Leggett}}, \bibinfo
  {author} {\bibfnamefont{S.}~\bibnamefont{Chakravarty}}, \bibinfo {author}
  {\bibfnamefont{A.~T.}\ \bibnamefont{Dorsey}}, \bibinfo {author}
  {\bibfnamefont{M.~P.~A.}\ \bibnamefont{Fisher}}, \bibinfo {author}
  {\bibfnamefont{A.}~\bibnamefont{Garg}},\ and\ \bibinfo {author}
  {\bibfnamefont{W.}~\bibnamefont{Zwerger}},\ }%
  \bibfield{journal}{%
  \Doi{10.1103/RevModPhys.59.1}{\bibinfo {journal} {Rev. Mod. Phys.}}\ }%
  \textbf{\bibinfo {volume} {59}},\ \bibinfo {pages} {1} (\bibinfo {year} {1987}).
\bibitem{hanggi19901}%
  \BibitemOpen
  \bibfield{author}{%
  \bibinfo {author} {\bibfnamefont{P.}~\bibnamefont{H\"anggi}}, \bibinfo
  {author} {\bibfnamefont{P.}~\bibnamefont{Talkner}},\ and\ \bibinfo {author}
  {\bibfnamefont{M.}~\bibnamefont{Borkovec}},\ }%
  \bibfield{journal}{%
  \Doi{10.1103/RevModPhys.62.251}{\bibinfo {journal} {Rev. Mod. Phys.}}\ }%
  \textbf{\bibinfo {volume} {62}},\ \bibinfo {pages} {251} (\bibinfo {year} {1990}).
  \bibitem{weiss1999}%
  \BibitemOpen
  \bibfield{author}{%
  \bibinfo {author} {\bibfnamefont{U.}~\bibnamefont{Weiss}},\ }%
  \emph{\bibinfo {title} {Quantum dissipative systems}},\ Vol.~\bibinfo
  {volume} {10}\ (\bibinfo {publisher} {World Scientific Publishing Company
  Incorporated},\ \bibinfo {year} {1999}).
\bibitem{unruh19891}%
  \BibitemOpen
  \bibfield{author}{%
  \bibinfo {author} {\bibfnamefont{W.~G.}\ \bibnamefont{Unruh}}\ and\ \bibinfo
  {author} {\bibfnamefont{W.~H.}\ \bibnamefont{Zurek}},\ }%
  \bibfield{journal}{%
  \Doi{10.1103/PhysRevD.40.1071}{\bibinfo {journal} {Phys. Rev. D}}\ }%
  \textbf{\bibinfo {volume} {40}},\ \bibinfo {pages} {1071} (\bibinfo {year} {1989}).
\bibitem{kohler20121}%
  \BibitemOpen
  \bibfield{author}{%
  \bibinfo {author} {\bibfnamefont{H.}~\bibnamefont{Kohler}}\ and\ \bibinfo
  {author} {\bibfnamefont{F.}~\bibnamefont{Sols}},\ }%
  \bibfield{journal}{%
  \Doi{10.1016/j.physa.2013.01.019}{\bibinfo {journal} {Physica A: Statistical
  Mechanics and its Applications}},\ }%
   (\bibinfo {year} {2013}),\ ISSN \bibinfo {issn} {0378-4371}.

\bibitem{hanggi20051}%
  \BibitemOpen
  \bibfield{author}{%
  \bibinfo {author} {\bibfnamefont{P.}~\bibnamefont{Hänggi}}\ and\ \bibinfo
  {author} {\bibfnamefont{G.-L.}\ \bibnamefont{Ingold}},\ }%
  \bibfield{journal}{%
  \bibinfo {journal} {Chaos}\ }%
  \textbf{\bibinfo {volume} {15}},\ \bibinfo {pages} {026105} (\bibinfo {year}
  {2005}).
  
\bibitem{giedke20011}%
  \BibitemOpen
  \bibfield{author}{%
  \bibinfo {author} {\bibfnamefont{G.}~\bibnamefont{Giedke}}, \bibinfo {author}
  {\bibfnamefont{B.}~\bibnamefont{Kraus}}, \bibinfo {author}
  {\bibfnamefont{M.}~\bibnamefont{Lewenstein}},\ and\ \bibinfo {author}
  {\bibfnamefont{J.~I.}\ \bibnamefont{Cirac}},\ }%
  \bibfield{journal}{%
  \Doi{10.1103/PhysRevA.64.052303}{\bibinfo {journal} {Phys. Rev. A}}\ }%
  \textbf{\bibinfo {volume} {64}},\ \bibinfo {pages} {052303} (\bibinfo {year} {2001}).
 
\bibitem{adesso20071}%
  \BibitemOpen
  \bibfield{author}{%
  \bibinfo {author} {\bibfnamefont{G.}~\bibnamefont{Adesso}}\ and\ \bibinfo
  {author} {\bibfnamefont{F.}~\bibnamefont{Illuminati}},\ }%
  \bibfield{journal}{%
  \bibinfo {journal} {J. Phy. A: Mathematical and Theoretical}\ }%
  \textbf{\bibinfo {volume} {40}},\ \bibinfo {pages} {7821} (\bibinfo {year}
  {2007}).  
 
\bibitem{hsiang20131}
J.-T. Hsiang, Rong Zhou, and B. L. Hu arXiv:1306.3728 (2013). 

\bibitem{valido20131}
  \BibitemOpen
  \bibfield{author}{%
  \bibinfo {author} {\bibfnamefont{A.~A.}\ \bibnamefont{Valido}}, \bibinfo
  {author} {\bibfnamefont{L.~A.}\ \bibnamefont{Correa}},\ and\ \bibinfo
  {author} {\bibfnamefont{D.}~\bibnamefont{Alonso}},\ }%
  \bibfield{journal}{%
  {\bibinfo {journal} {Phys. Rev. A}}\ }%
  \textbf{\bibinfo {volume} {88}},\ \bibinfo {pages} {012309} (\bibinfo {year} {2013}).

\bibitem{coleman19621}%
  \BibitemOpen
  \bibfield{author}{%
  \bibinfo {author} {\bibfnamefont{S.}~\bibnamefont{Coleman}}\ and\ \bibinfo
  {author} {\bibfnamefont{R.~E.}\ \bibnamefont{Norton}},\ }%
  \bibfield{journal}{%
  \Doi{10.1103/PhysRev.125.1422}{\bibinfo {journal} {Phys. Rev.}}\ }%
  \textbf{\bibinfo {volume} {125}},\ \bibinfo {pages} {1422} (\bibinfo {year} {1962}).
\bibitem{aitken1956}%
  \BibitemOpen
  \bibfield{author}{%
  \bibinfo {author} {\bibfnamefont{A.~C.}\ \bibnamefont{Aitken}}\ and\ \bibinfo
  {author} {\bibfnamefont{A.~C.}\ \bibnamefont{Aitken}},\ }%
  \emph{\bibinfo {title} {Determinants and matrices}},\ Vol.~\bibinfo {volume}
  {1}\ (\bibinfo {publisher} {Oliver and Boyd Edinburgh and London},\ \bibinfo
  {year} {1956}).
\bibitem{zhang20121}%
  \BibitemOpen
  \bibfield{author}{%
  \bibinfo {author} {\bibfnamefont{W.-M.}\ \bibnamefont{Zhang}}, \bibinfo
  {author} {\bibfnamefont{P.-Y.}\ \bibnamefont{Lo}}, \bibinfo {author}
  {\bibfnamefont{H.-N.}\ \bibnamefont{Xiong}}, \bibinfo {author}
  {\bibfnamefont{MW.-Y.}\ \bibnamefont{Tu}},\ and\ \bibinfo {author}
  {\bibfnamefont{F.}~\bibnamefont{Nori}},\ }%
  \bibfield{journal}{%
  \Doi{10.1103/PhysRevLett.109.170402}{\bibinfo {journal} {Phys. Rev. Lett.}}\
  }%
  \textbf{\bibinfo {volume} {109}},\ \bibinfo {pages} {170402} (\bibinfo {year} {2012}).

\bibitem{vidal20021}%
  \BibitemOpen
  \bibfield{author}{%
  \bibinfo {author} {\bibfnamefont{G.}~\bibnamefont{Vidal}}\ and\ \bibinfo
  {author} {\bibfnamefont{R.~F.}\ \bibnamefont{Werner}},\ }%
  \bibfield{journal}{%
  \Doi{10.1103/PhysRevA.65.032314}{\bibinfo {journal} {Phys. Rev. A}}\ }%
  \textbf{\bibinfo {volume} {65}},\ \bibinfo {pages} {032314} (\bibinfo {year} {2002}).
\bibitem{marian20121}%
  \BibitemOpen
  \bibfield{author}{%
  \bibinfo {author} {\bibfnamefont{P.}~\bibnamefont{Marian}}\ and\ \bibinfo
  {author} {\bibfnamefont{T.~A.}\ \bibnamefont{Marian}},\ }%
  \bibfield{journal}{%
  \Doi{10.1103/PhysRevA.86.022340}{\bibinfo {journal} {Phys. Rev. A}}\ }%
  \textbf{\bibinfo {volume} {86}},\ \bibinfo {pages} {022340} (\bibinfo {year} {2012}).
\bibitem{zhang20071}%
  \BibitemOpen
  \bibfield{author}{%
  \bibinfo {author} {\bibfnamefont{J.-H.}\ \bibnamefont{An}}\ and\ \bibinfo
  {author} {\bibfnamefont{W.-M.}\ \bibnamefont{Zhang}},\ }%
  \bibfield{journal}{%
  \Doi{10.1103/PhysRevA.76.042127}{\bibinfo {journal} {Phys. Rev. A}}\ }%
  \textbf{\bibinfo {volume} {76}},\ \bibinfo {pages} {042127} (\bibinfo {year} {2007}).
\bibitem{ludwig20101}%
  \BibitemOpen
  \bibfield{author}{%
  \bibinfo {author} {\bibfnamefont{M.}~\bibnamefont{Ludwig}}, \bibinfo {author}
  {\bibfnamefont{K.}~\bibnamefont{Hammerer}},\ and\ \bibinfo {author}
  {\bibfnamefont{F.}~\bibnamefont{Marquardt}},\ }%
  \bibfield{journal}{%
  \Doi{10.1103/PhysRevA.82.012333}{\bibinfo {journal} {Phys. Rev. A}}\ }%
  \textbf{\bibinfo {volume} {82}},\ \bibinfo {pages} {012333} (\bibinfo {year} {2010}).
\bibitem{riseborough19851}%
  \BibitemOpen
  \bibfield{author}{%
  \bibinfo {author} {\bibfnamefont{P.~S.}\ \bibnamefont{Riseborough}}, \bibinfo
  {author} {\bibfnamefont{P.}~\bibnamefont{Hanggi}},\ and\ \bibinfo {author}
  {\bibfnamefont{U.}~\bibnamefont{Weiss}},\ }%
  \bibfield{journal}{%
  \Doi{10.1103/PhysRevA.31.471}{\bibinfo {journal} {Phys. Rev. A}}\ }%
  \textbf{\bibinfo {volume} {31}},\ \bibinfo {pages} {471} (\bibinfo {year} {1985}).
\bibitem{benatti20111}%
  \BibitemOpen
  \bibfield{author}{%
  \bibinfo {author} {\bibfnamefont{F.}~\bibnamefont{Benatti}}\ and\ \bibinfo
  {author} {\bibfnamefont{A.}~\bibnamefont{Nagy}},\ }%
  \bibfield{journal}{%
  \Doi{10.1016/j.aop.2010.09.006}{\bibinfo {journal} {Ann. Phys.}}\ }%
  \textbf{\bibinfo {volume} {326}},\ \bibinfo {pages} {740 } (\bibinfo {year}
  {2011}),\ ISSN \bibinfo {ISSN} {0003-4916}.
\bibitem{an20111}%
  \BibitemOpen
  \bibfield{author}{%
  \bibinfo {author} {\bibfnamefont{N.~B.}\ \bibnamefont{An}}, \bibinfo {author}
  {\bibfnamefont{J.}~\bibnamefont{Kim}},\ and\ \bibinfo {author}
  {\bibfnamefont{K.}~\bibnamefont{Kim}},\ }%
  \bibfield{journal}{%
  \Doi{10.1103/PhysRevA.84.022329}{\bibinfo {journal} {Phys. Rev. A}}\ }%
  \textbf{\bibinfo {volume} {84}},\ \bibinfo {pages} {022329} (\bibinfo {year} {2011}).
  
\bibitem{anders20081}%
  \BibitemOpen
  \bibfield{author}{%
  \bibinfo {author} {\bibfnamefont{J.}~\bibnamefont{Anders}},\ }%
  \bibfield{journal}{%
  \Doi{10.1103/PhysRevA.77.062102}{\bibinfo {journal} {Phys. Rev. A}}\ }%
  \textbf{\bibinfo {volume} {77}},\ \bibinfo {pages} {062102} (\bibinfo {year} {2008}). 
\bibitem{bennet20111}%
  \BibitemOpen
  \bibfield{author}{%
  \bibinfo {author} {\bibfnamefont{C.~H.}\ \bibnamefont{Bennett}}, \bibinfo
  {author} {\bibfnamefont{A.}~\bibnamefont{Grudka}}, \bibinfo {author}
  {\bibfnamefont{M.}~\bibnamefont{Horodecki}}, \bibinfo {author}
  {\bibfnamefont{P.}~\bibnamefont{Horodecki}},\ and\ \bibinfo {author}
  {\bibfnamefont{R.}~\bibnamefont{Horodecki}},\ }%
  \bibfield{journal}{%
  \Doi{10.1103/PhysRevA.83.012312}{\bibinfo {journal} {Phys. Rev. A}}\ }%
  \textbf{\bibinfo {volume} {83}},\ \bibinfo {pages} {012312} (\bibinfo {year} {2011}).
\bibitem{erdelyi1954}%
  \BibitemOpen
  \bibfield{author}{%
  \bibinfo {author} {\bibfnamefont{A.}~\bibnamefont{Erd{\'e}lyi}}, \bibinfo
  {author} {\bibfnamefont{W.}~\bibnamefont{Magnus}}, \bibinfo {author}
  {\bibfnamefont{F.}~\bibnamefont{Oberhettinger}},\ and\ \bibinfo {author}
  {\bibfnamefont{F.~G.}\ \bibnamefont{Tricomi}},\ }%
  \emph{\bibinfo {title} {Tables of Integral Transforms: Vol.: 2}}\ (\bibinfo
  {publisher} {McGraw-Hill Book Company, Incorporated},\ \bibinfo {year}
  {1954}).
 
  \bibitem{serafini20051}%
  \BibitemOpen
  \bibfield{author}{%
  \bibinfo {author} {\bibfnamefont{A.}~\bibnamefont{Serafini}}, \bibinfo
  {author} {\bibfnamefont{G.}~\bibnamefont{Adesso}},\ and\ \bibinfo {author}
  {\bibfnamefont{F.}~\bibnamefont{Illuminati}},\ }%
  \bibfield{journal}{%
  \Doi{10.1103/PhysRevA.71.032349}{\bibinfo {journal} {Phys. Rev. A}}\ }%
  \textbf{\bibinfo {volume} {71}},\ \bibinfo {pages} {032349} (\bibinfo {year} {2005}).
\end{thebibliography}
%

\end{document}